%% file: nloimpact.tex
\documentclass[11pt, a4paper]{article} 

\input{wenneker.tex} 
\graphicspath{{figs/}}

\renewcommand\Im{\operatorname{\mathfrak{Im}}}

\newcommand{\Ht}{\widetilde{\mathcal{H}}}

\newcommand{\ImH}{\Im{\mathcal{H}}}

\newcommand{\PG}{\HepParticle{G}{}{}\xspace}
\newcommand{\Psea}{\HepParticle{sea}{}{}\xspace}
\newcommand{\PgammastarL}{\HepParticle{\gamma}{L}{*}\xspace}
\newcommand{\PgammastarT}{\HepParticle{\gamma}{T}{*}\xspace}
\newcommand{\PVL}{\HepParticle{V}{L}{}\xspace}

\newcommand{\PVLO}{\HepParticle{V}{L}{0}\xspace}

\newcommand{\DVCS}{\HepProcess{\Pgammastar \Pproton \to \Pgamma \Pproton}\xspace}
\newcommand{\DVMPrho}{\HepProcess{\Pgammastar \Pproton \to \Prhozero \Pproton}\xspace}
\newcommand{\DVMPphi}{\HepProcess{\Pgammastar \Pproton \to \Pphi \Pproton}\xspace}

\newcommand{\dvvlp}{{\rm DV}\PVL{\rm P}\xspace}
\newcommand{\dvvlpO}{{\rm DV}\PVLO{\rm P}\xspace}
\newcommand{\dvcs}{{\rm DVCS}}

\newcommand{\sigDVCS}{\sigma^{\Pgamma}}
\newcommand{\sigDVMP}{\sigma^{\Prho^0_{\rm L}}} 
\newcommand{\sigDVMPL}{\sigma^{\Prho^0}_{\textnormal{L}}}
\newcommand{\sigDVMPT}{\sigma^{\Prho^0}_{\textnormal{T}}}

\newcommand{\xb}{x_{\mathrm{B}}}
\newcommand{\Q}{\mathcal{Q}}
\newcommand{\chinpts}{\chi^2/n_{\textnormal{pts}}}
\newcommand{\mrho}{m_{\Prho^0}}
\newcommand{\alphas}{\alpha_{\textnormal{s}}}
\newcommand{\msbar}{\overline{\textnormal{MS}}}
\renewcommand{\d}{d} 

\newcommand{\model}[1]{\texttt{#1}}  

\newcommand{\req}[1]{(\ref{#1})}

\newcommand{\muR}{\mu_{\rm R}}

\newcommand{\muGPD}{\mu_{\rm GPD}}
\newcommand{\muDA}{\mu_{\rm DA}}
\newcommand{\muO}{\mu_0}
\newcommand{\Tevol}{\mbox{\boldmath $\bar{T}$}}
\newcommand{\tevol}{\mbox{$\bar{T}$}}

\newcommand{\bEop}{\mathbb{E}} 
\newcommand{\Eop}{{\rm E}} 

\newcommand{\ootimes}{\widetilde{\otimes}} 

\title{\begin{center}\Huge NLO corrections to the deeply virtual meson production \\ 
revisited: impact on the extraction of generalized parton distributions\end{center}}

\author{
    \authorstyle{Marija Čuić\textsuperscript{1}},
    \authorstyle{Goran Duplančić\textsuperscript{2}},
    \authorstyle{Krešimir Kumerički\textsuperscript{1}},
    \authorstyle{Kornelija Passek-K.\textsuperscript{2}} 
	\newline\newline 
	\textsuperscript{1}\institution{Department of Physics, Faculty of Science, University of Zagreb,
    HR-10000 Zagreb, Croatia}\\ 
    \textsuperscript{2}\institution{Division of Theoretical Physics, Ruđer Bošković Institute,
    HR-10002 Zagreb, Croatia}\\ 
}

\date{\today}

\begin{document}

\maketitle 

\thispagestyle{firstpage} 


\lettrineabstract{%
We revisit the next-to-leading order (NLO) perturbative QCD corrections for the
deeply virtual meson production (DVMP) process, exploring its phenomenology
both in isolation and in a multichannel fit combined with deeply virtual
Compton scattering (DVCS). Our approach involves the conformal partial wave
(CPaW) formalism, which allows for the straightforward inclusion of
higher-order contributions and evolutionary effects.
Our findings indicate that a description of the longitudinal component of the
vector meson DVMP cross-section at high energies is achievable only at NLO
within the standard collinear approach. Furthermore, we demonstrate a simultaneous
description of DIS, DVCS, and DVMP processes, providing insights into the
proton structure described at NLO by unique universal generalized parton
distribution (GPD) functions.
}

\clearpage

\tableofcontents


\section{Introduction}

In the family of structure functions describing the quark-gluon structure
of hadrons, generalized parton distributions (GPDs)
\cite{Mueller:1998fv,Radyushkin:1996nd,Ji:1996nm}
occupy a special place.
Their understanding would provide insights into several tangible physical
properties of hadrons, such as the composition of their spin \cite{Ji:1996ek},
3D tomography \cite{Burkardt:2000za}, or pressure \cite{Polyakov:2002yz}.
Simultaneously, the processes enabling their experimental determination are
measurable in both current and forthcoming experimental facilities
\cite{Accardi:2012qut,Anderle:2021wcy}.
Processes that in this sense are the subject of the most attention of the
scientific community are deeply virtual Compton scattering (DVCS) and deeply
virtual meson production (DVMP), for whose phenomenological status see
\cite{dHose:2016mda,Kumericki:2016ehc,Favart:2015umi} and references therein.
Other processes 
have also been identified as possibly important
sources of information about GPDs, such as
timelike DVCS \cite{Berger:2001xd,
Pire:2011st,
Boer:2015fwa,
Grocholski:2019pqj},
double DVCS \cite{Belitsky:2002tf,Guidal:2002kt,Pire:2011st,Deja:2023ahc}, or
boson pair production 
\cite{Ivanov:2002jj,Boussarie:2016qop,Pedrak:2017cpp,%
Siddikov:2022bku,
Duplancic:2022ffo},
see also 
\cite{Qiu:2022pla}%
, 
but experimental measurements of these are so far scarce or nonexistent%
\footnote{The list is not exhaustive 
and we refer to, for example \cite{Diehl:2003ny},
for the account on processes 
providing complementary
information on GPDs
from exclusive processes with
a large momentum transfer to a nucleon
(large $t$) such as 
nucleon form factors
\cite{Diehl:2013xca}, 
wide-angle Compton scattering
\cite{Radyushkin:1998rt,Diehl:1998kh}
and wide-angle meson production
\cite{Huang:2000kd,Kroll:2021ecb},
as well as crossed processes.
Their analysis is experimentally
and theoretically more 
challenging.}%
.
So in this paper, we %
focus on DVCS and DVMP,
in particular, the production of longitudinally polarized 
vector mesons (\dvvlp). 
Notably, transversely polarized vector meson production vanishes
at leading-twist \cite{Collins:1999un}. 
Moreover, at experimentally accessible energies, pseudoscalar meson production 
appears to be predominantly influenced by higher-twist effects 
\cite{Goloskokov:2009ia, Goloskokov:2011rd}.

The significance and complementary nature of DVCS and DVMP 
in unveiling GPDs have long been recognized. 
This is evident in the ongoing efforts to enhance the precision 
of their theoretical descriptions 
by calculating higher-order perturbative QCD (pQCD) corrections
to subprocess hard-scattering amplitudes:
NLO DVCS \cite{Ji:1997nk, Belitsky:1997rh, Mankiewicz:1997bk, Ji:1998xh},
NLO DVMP \cite{Belitsky:2001nq, Ivanov:2004zv, 
Duplancic:2016bge},
and
NNLO DVCS \cite{Braun:2020yib, Braun:2021grd, Ji:2023xzk}.
Additionally, the power-corrections to DVCS have been determined
\cite{Braun:2011zr, Braun:2011dg, Braun:2014sta, Guo:2021gru, Braun:2022qly},
while the consistent inclusion of power-corrections in DVMP remains a challenging endeavor.
The leading twist-2 contribution
to DVMP
accounts for the contribution 
of longitudinally polarized photons
(\HepProcess{\PgammastarL \PN \to \HepParticle{M}{}{} \PN'}).
Currently, there remains a challenge in systematically distinguishing 
between experimental data associated with longitudinal and transverse contributions.
Conversely, for DVCS the leading twist pertains to the contribution of
transverse photons 
(\HepProcess{\PgammastarT  \PN \to \Pgamma \PN}).

Unfortunately, the application of the existing theoretical apparatus to
phenomenology and to the extraction of GPDs is currently not at a satisfactory
level. Analyses most often focus on only one of these processes, and even then
most often they remain at the level of the leading order (LO) of the
perturbation theory.  
Bearing in mind that in the most studied DVCS process the
gluon contribution appears at NLO, it is clear that the resulting
extractions of GPDs are not reliable.  
The researchers themselves are mostly aware of this; 
therefore%
, the largest number of previous studies stops at the
determination of Compton form factors (CFF) that combine GPDs of all 
flavors
and corresponding hard scattering amplitudes.

Notable earlier attempts to grasp GPDs by simultaneous analysis 
of DVMP and DVCS processes are:
\begin{itemize}
     \item Ref. \cite{Kroll:2012sm}, where it was shown that the GPDs obtained from 
an earlier fit to DVMP data \cite{Goloskokov:2005sd,Goloskokov:2007nt,Goloskokov:2009ia}
also relatively 
reasonably describe the observables of DVCS, in particular those from the HERA collider. 
The approach used relies on inclusion of transversal degrees of freedom in a way for which 
it is not clear how to naturally extend it to higher orders of perturbative QCD expansion.
     \item Refs. \cite{Meskauskas:2011aa,Lautenschlager:2013uya,Lautenschlager:2015mxt},
where it has been shown that a relatively acceptable simultaneous fit of GPDs 
to DVV${}^{0}$P and DVCS data obtained from the HERA collider at LO and NLO is possible.
\end{itemize}

The continuous increase in the volume and precision of available experimental measurements, 
along with the promising capabilities of future research facilities, 
necessitates the development of a robust theoretical framework. 
This framework should be on par with those established in related fields, 
such as the extraction of ordinary parton distribution functions (PDFs)
and transverse momentum distributions (TMDs), extending at least to the NLO level, 
and potentially advancing to the NNLO level in the future.
The foundation of such a framework, which is based 
on the conformal partial wave (CPaW) 
formalism
\cite{Mueller:2005ed,Mueller:2005nz}
--- specifically, the conformal partial wave expansion in conjunction 
with the Mellin-Barnes integral technique
--- was established and systematically organized for DVCS and DVMP 
in \cite{Kumericki:2007sa,Mueller:2013caa}. 
In comparison to the traditional momentum fraction representation, 
it
not only  offers easier inclusion of GPD evolution 
to NLO order and beyond, but also
opens the door to intriguing GPD modeling possibilities 
and paves the way for the development of stable and efficient computer 
codes capable of handling GPD evolution and fitting 
to both experimental and lattice data.

The comprehensive list of NLO contributions to deeply virtual
production of vector (scalar) mesons
in both the conformal moment and momentum fraction representations 
has been systematically organized in \cite{Mueller:2013caa}. 
Similarly, the corresponding expressions for pseudoscalar (axial-vector) meson production, 
were derived in \cite{Duplancic:2016bge}, 
where additionally an omission in the NLO expressions for \dvvlpO was recognized. 
Here we present the corrected NLO expressions for \dvvlpO in the conformal moment representation.

The main motivation of this paper is to continue the application of CPaW formalism
to NLO analysis of DVMP and 
extraction of GPDs simultaneously 
from DVMP and DVCS data, for the first time systematically at the NLO level.
The only predecessor is the unpublished work \cite{Lautenschlager:2013uya},
with respect to which we bring here the following improvements:
\begin{itemize}
     \item The expressions for the NLO hard-scattering \dvvlpO amplitude 
have been corrected in the meantime \cite{Duplancic:2016bge}.
     \item 
Here we treat the measured cross sections as a reliable experimental result.
     In paper \cite{Lautenschlager:2013uya} the normalizations 
of the experimentally measured cross sections for \dvvlpO were treated 
as fitting parameters. 
     \item The GPD model we use here is of the same type 
as in \cite{Lautenschlager:2013uya}, but simpler, with fewer free parameters.
\end{itemize}

In addition to the improved extraction of GPDs, the aim of this work is to
study more carefully the impact of NLO corrections.  We find that despite the
very different roles of quarks and gluons in DVCS and
\dvvlpO, both processes at
the NLO level give very similar GPDs, thus pointing out at the same time the universality
of GPDs and the unquestionable need for a NLO theoretical approach.

In previous papers mostly dedicated to the calculation of NLO
corrections, and not so much to applications, the impact of NLO corrections is usually studied in a simplified
environment, using some fixed toy model, the same for LO and NLO.
With such an approach it is possible only to get some idea about the size of the coefficients in a given order of
pQCD expansion 
(see, for example, detailed analysis in \cite{Kumericki:2007sa,Mueller:2013caa}). 
Here, 
we want to answer a complementary, phenomenologically more
important question: How does the shape of the GPD change when going from LO to NLO?
In this sense, we exactly follow the proposition of the authors of \cite{Mueller:2013caa}
who consider
their analysis of the impact of NLO corrections ``not entirely realistic''
and suggest that ``appropriate method [$\ldots$] would be the quantification 
of reparametrization effects that arise from LO and NLO fits to experimental data.''

To make our main message, that universal NLO GPDs can describe DVMP and DVCS
processes at high energies (small Bjorken $\xb$) in the collinear pQCD approach, as
clear as possible and free from technical complications, we assume that
the only contribution to the amplitudes comes from the dominant sea quark and
gluon GPDs $H^{\Psea}$ and $H^{\PG}$, which is a good approximation in our
chosen kinematics.
Also, the only observables we consider are the cross sections for the processes
\DVCS (DVCS) and \DVMPrho (DVMP).
Other observables or lattice data, in other kinematic regimes, would require 
the addition of valence quarks, other GPDs, and the improvement of our GPD model, 
which we leave for future works.

In Section \ref{sec:pqcd} we outline the main elements 
of collinear perturbative QCD framework and specifically of CPaW formalism.
We summarize the LO and NLO contributions to subprocess hard-scattering 
amplitudes for \dvvlp and DVCS, and explain the implementation of GPD evolution to NLO.
In Section \ref{sec:gpdmodel} we specify the GPD model we use, and
in Section \ref{sec:exp} the used experimental measurements are presented, with special
emphasis on the problem of separation of the longitudinal and transverse
components of the cross section for meson production.
In Section \ref{sec:results}, we present and discuss our results,
while the concluding remarks are given in Section \ref{sec:concl}.

\section{Perturbative QCD framework}
\label{sec:pqcd}

\begin{figure}[t]
    \centering{\includegraphics[width=0.45\linewidth]{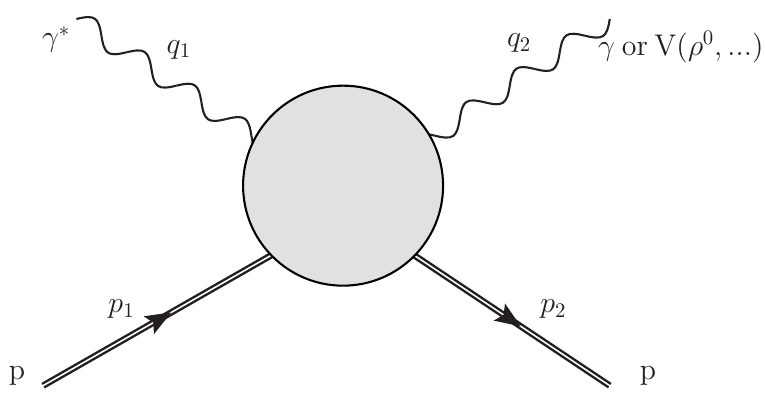}}
    \caption{Kinematics of the DVCS and DVMP processes.}
    \label{fig:kinematics}
\end{figure}

In this paper, we use the standard collinear perturbative QCD approach
\cite{Lepage:1980fj,Efremov:1979qk} for both observed processes,
DVCS and DVMP, specifically \dvvlp, Fig. \ref{fig:kinematics}.
This is worth emphasizing considering that the first attempts to apply this
approach to DVMP \cite{Brodsky:1994kf,Martin:1996bp} gave results for
cross sections in discrepancy with measurements. 
The reason for this is, among other things, the effective $\sim 1/\Q^4$
scaling of measured cross sections, visible in Fig. \ref{fig:Q2scaling},
inconsistent with the canonical asymptotic $\sim 1/\Q^6$ scaling that follows
from the collinear pQCD amplitude. In the currently most popular approach to the calculation
of DVMP cross sections within GPD framework \cite{Goloskokov:2005sd,Goloskokov:2007nt}, this is
fixed by introducing power corrections due to transverse momentum
of partons. This approach has proved to be quite successful phenomenologically,
but it is not clear how to apply it consistently at NLO.
As we confirm in this paper, 
NLO corrections are significant, and the authors
of the early studies of these corrections \cite{Diehl:2007hd} %
concluded that
for quantitative control in the high-energy regime, 
it is necessary to perform a resummation of large BFKL
type $\log({1}/{\xb})$ logarithms \cite{Ivanov:2007je}.  
In the point of view to which we adhere \cite{Mueller:2013caa}, 
the change of
effective scaling is achieved by $\log\Q^2$ perturbative logarithms.

\begin{figure}[t]
    \centering{\includegraphics[width=0.8\linewidth]{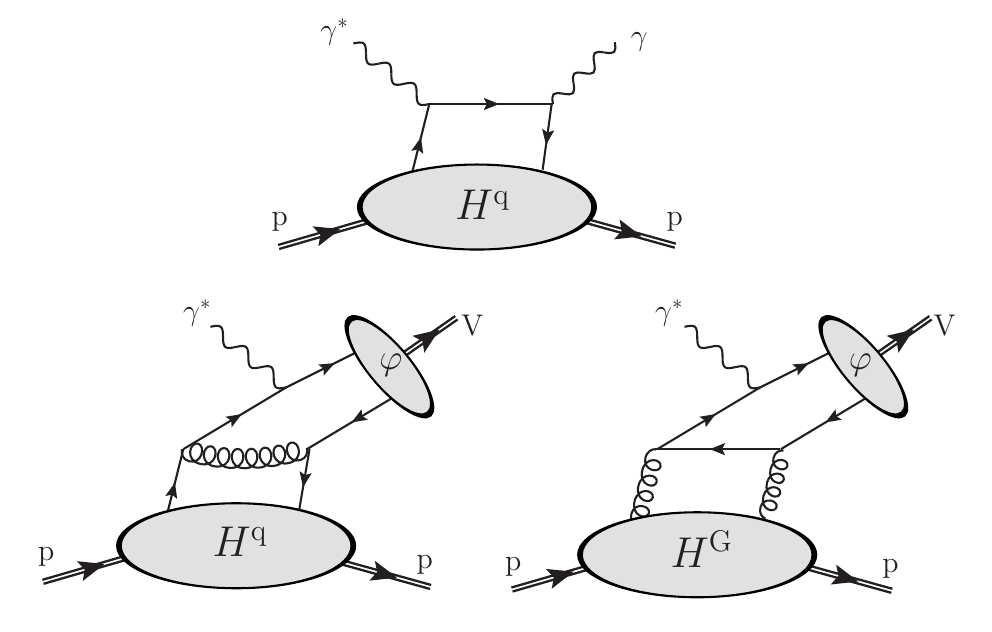}}
    \caption{Generic diagrams contributing to the factorized DVCS (top) and DVMP (bottom) process at LO.
     In the DVMP case, the gluon GPD at input scale contributes already at LO (lower right).}
    \label{fig:DVMPDVCS}
\end{figure}

The collinear perturbative QCD approach relies on the property of factorization
of the amplitude into the non-perturbative
soft structure functions 
and the subprocess amplitude of hard scattering on the active parton
\cite{Collins:1996fb,Collins:1998be}, see
Fig. \ref{fig:DVMPDVCS}.
This approach
provides the leading-twist contributions 
that describe 
the
transversal DVCS (\HepProcess{\PgammastarT \, \Pproton \to \Pgamma \Pproton}) 
and 
longitudinal \dvvlp (\HepProcess{\PgammastarL \, \Pproton \to \PVL \, \Pproton}) 
cross section components.
In the following, we present
the main points of this approach,
taking the opportunity to collect and summarize 
the relevant expressions, with a particular focus 
on the CPaW formalism. 
Additional details can be found in 
\cite{Kumericki:2007sa,Mueller:2013caa,Duplancic:2016bge}.

The observables are expressed using process amplitudes 
termed Compton form factors (CFFs) for DVCS and 
transition form factors (TFFs) for DVMP. 
The nomenclature aligns with 
the nomenclature used for the contributing GPDs.
Additionally, TFFs depend on meson distribution amplitudes (DAs), 
representing the meson's internal structure,
making the analysis
of the process more challenging, 
but also potentially more rewarding.

In the case of DVCS at twist-2, contributions arise 
from intrinsic parity even (vector) GPDs denoted as $H^a$ and $E^a$, 
as well as intrinsic parity odd (axial-vector) GPDs 
denoted as $\widetilde{H}^a$ and $\widetilde{E}^a$\footnote{In this work
we do not consider chiral-odd transversity GPDs.}.
Here, the variable $a$ ranges over different quark flavors (\Pqu, \Pqd, \Pqs, \ldots) 
and gluon contributions ($\PG$). 
The gluon contributions to the subprocess hard-scattering amplitude 
of DVCS emerge at NLO. However, their impact on CFFs at LO arises 
through GPD evolution, attributed to the mixing of quark singlet 
and gluon GPDs.

In contrast, DVMP provides a natural distinction of GPDs of different parity. 
For the specific case of vector meson production (\dvvlp), 
only the intrinsic parity even (vector) GPDs $H^a$ and $E^a$ contribute. 
On the other hand, the GPDs $\widetilde{H}^q$ and $\widetilde{E}^q$ 
come into play in the production of pseudoscalar mesons
where there are no GPD gluon contributions%
\footnote{For details on all meson channels, refer to table (2.30) in \cite{Mueller:2013caa}.}. 
Notably, in the production of neutral $\PVL$ mesons, 
the contributions of gluon GPDs $H^{\PG}$ and $E^{\PG}$ are more significant, 
as, unlike DVCS,  these gluon GPDs already contribute 
to the LO subprocess hard-scattering amplitude (see Fig. \ref{fig:DVMPDVCS}).

In the twist-2 approximation 
and for the small $\xb$ (regime 
of current interest in phenomenological analysis) 
the differential cross section for DVCS 
(\HepProcess{\PgammastarT \Pproton \to \Pgamma \Pproton}) is given by
\begin{equation}
    \frac{\d \sigma^{\Pgamma}}{\d t} = \pi \alpha_{\rm em}^2 \frac{\xb^2}{\Q^4}  \bigg\{
        \left| \mathcal{H}(\xb, t, \Q^2) \right|^2 - \frac{t}{4 M_{\Pproton}^2}
        \left| \mathcal{E}(\xb, t, \Q^2) \right|^2 +
        \left| \Ht(\xb, t, \Q^2) \right|^2 \bigg\} \,,
    \label{eq:sigDVCSt}
\end{equation}
where $\alpha_{\rm em}$ is the electromagnetic fine structure constant, $M_{\Pproton}$ is the
proton mass, Mandelstam $t = (p_{1}-p_{2})^2$, $\Q^2 = - q_{1}^2$ is the virtuality
of the initial photon, and $\xb = {\Q^2}/{(2 p_1 \cdot q_1) }$ is the Bjorken variable.
Momenta $p_{1,2}$ and $q_{1,2}$ are specified on Fig.~\ref{fig:kinematics}.
In this approximation three, out of the four, CFFs contribute, 
$\mathcal{H}$, $\mathcal{E}$ and $\Ht$,
and we neglect terms suppressed by further powers of small $\xb$.
For small $\xb$ arguments from Regge theory suggest  $\mathcal{H} \sim {1}/{\xb}$
and $\Ht \sim {1}/\sqrt{\xb}$, so it is justified to neglect the
contribution of $\Ht$. The situation with $\mathcal{E}$ is less clear, but
we still rely on the suppression factor ${t}/{(4 M_{\Pproton}^2)}$ and neglect
$\mathcal{E}$ in 
the numerical treatment.

Similarly, at twist-2 the differential cross section for 
\dvvlp (\HepProcess{\PgammastarL \,\Pproton \to \PVL \,\Pproton})
in the small $\xb$ regime simplifies to
\begin{equation}
    \frac{\d \sigma^{\PVL}}{\d t} =  4 \pi^2 \alpha_{\rm em} \frac{\xb^2}{\Q^4}  \bigg\{
        \left| \mathcal{H}_{\PVL}(\xb, t, \Q^2) \right|^2 - \frac{t}{4 M_{\Pproton}^2}
        \left| \mathcal{E}_{\PVL}(\xb, t, \Q^2) \right|^2
         \bigg\} \,,
    \label{eq:sigDVMPt}
\end{equation}
where $\mathcal{H}_{\PVL}$ and $\mathcal{E}_{\PVL}$ are TFFs
(\HepProcess{\Pproton \to \PVL \, \Pproton}) for the light vector meson \PVL.
For the same reason as in the DVCS case in this work
we ignore the contribution of $\mathcal{E}_{\PVL}$.

\subsection{Factorization of form factors and CPaW formalism}

As argued above, considering the specific energy regime of interest
in this study, the differential cross sections are
expressed solely in terms of the Compton form factor 
$\mathcal{H}$ for DVCS 
and the transition form factor $\mathcal{H}_{\PVL}$ for \dvvlp. 
In the following we illustrate the factorization of form factors 
using these two as examples.
The given expressions remain valid across all energy regimes 
and are applicable to other form factors,
in particular to $\mathcal{E}$ and $\mathcal{E}_{\PVL}$, 
taking care to use the subprocess hard-scattering amplitudes suitable 
for the intrinsic parity involved.

Both CFF $\mathcal{H}$ and TFF $\mathcal{H}_{\PVL}$,
according to the factorization property, can be written as a convolution of the corresponding
subprocess hard-scattering amplitudes $T_{\dvcs}$ and $T_{\dvvlp}$ and
non-perturbative soft functions, GPD $H$ and, in the case of $\dvvlp$,
 the meson distribution amplitude
$\varphi_{\PVL}$:
\begin{eqnarray}
    \mathcal{H}(\xb, t, \Q^2) 
& = &  
H^a(x, \xi, t, \muGPD^2)
\:\stackrel{x}{\otimes}\: 
{}^aT^{\rm DVCS}\left(\!x, \xi, \alphas(\muR),
        \frac{\Q^2}{\muR^2}, \frac{\Q^2}{\muGPD^2} \right)\,,
                       \label{eq:Hconvolution} 
\end{eqnarray}

\begin{eqnarray}
\lefteqn{\mathcal{H}_{\PVL}(\xb, t, \Q^2) }
\nonumber \\
&=& \frac{C_{\rm F} f_{\PVL}}{N_c \Q}
\:
H^a(x, \xi, t, \muGPD^2)  
\:\stackrel{x}{\otimes}\: 
{}^aT^{\dvvlp}\left(\!x, \xi, v, \alphas(\muR),
\frac{\Q^2}{\muR^2}, \frac{\Q^2}{\muGPD^2}, 
\frac{\Q^2}{\muDA^2} \right)
\:\stackrel{v}{\otimes}\:
 \varphi_{\PVL}(v, \muDA^2) 
\:
\, ,
\nonumber \\
              \label{eq:HVconvolution}
\end{eqnarray}
where
\begin{equation}
    \xi = \frac{\xb}{2-\xb} \,,
    \label{eq:xixb}
\end{equation}
$C_{\rm F} = {4}/{3}$, $N_{\rm c} = 3$, DA $\varphi_{\PVL}$ is
normalized by extracting the meson decay constant $f_{\PVL}$ so that
\begin{equation}
    \int_{0}^{1} \d v \: \varphi_{\PVL}(v, \muDA^2) = 1 \,,
    \label{eq:DAnormalization}
\end{equation}
and $\stackrel{x}{\otimes}$ and $\stackrel{v}{\otimes}$ are 
defined in (\ref{eq:convolution}) below.
The GPDs are functions of three variables: 
$x$ --- the parton's ''average'' longitudinal momentum fraction,
$\xi$ --- the longitudinal momentum transfer (skewness parameter), 
and $t$ --- the momentum transfer squared,
while their evolution with energy is encapsulated in
the dependence on the factorization scale $\muGPD$. 
For the deeply virtual processes under consideration, 
skewness relates to $x_B$ through \req{eq:xixb}.
The hard-scattering amplitude at twist-2
depends on $x$, $\xi$, $\Q^2$, renormalization and factorization scales
and, in the DVMP case, $v$ --- the longitudinal momentum fraction of 
the leading Fock state parton in a meson%
\footnote{
In the collinear approximation, the momentum fractions of the emitted 
and reabsorbed partons are $u=(\xi+x)/(2 \xi)$ and $\bar{u}=1-u$, 
respectively. Frequently, hard-scattering amplitudes are expressed 
in this notation \cite{Mueller:2013caa}. 
The appropriate ''$+i \epsilon$'' prescription in DVCS and DVMP kinematics 
is restored by employing $\xi \to \xi -i \epsilon$.
}. 
In this collinear framework, the dependence of CFFs and TFFs 
on momentum transfer $t$ solely arises from GPDs.

Formally, the all-order predictions of factorized expressions 
(\ref{eq:Hconvolution}) and (\ref{eq:HVconvolution}) 
are scale-independent with respect to renormalization ($\muR$) 
and factorization ($\muGPD$, $\muDA$) scales. 
However, finite-order predictions do exhibit residual scale dependence.
Notably, 
the renormalization scale dependence is particularly pronounced in DVMP, 
as it comes into play already at LO 
in contrast to the DVCS where it appears at NLO
(see Fig. \ref{fig:DVMPDVCS}). 
Consequently, the stabilization effect of higher-order corrections starts 
at NLO for DVMP and at NNLO for DVCS. 
In both processes, gluon contributions are proportional to $\alphas$ 
and therefore sensitive to the choice of $\muR$.
The scale settings
are discussed in more detail in \cite{Kumericki:2007sa,Mueller:2013caa}.
It's worth noting that there are proposals in the literature 
regarding the 'optimal' scale settings 
and/or different resummation of factorization logarithms. 
However, these aspects are left for future investigation. 
For numerical evaluations in this study, we set the renormalization scale 
and both factorization scales to be equal to the photon virtuality, 
denoted as $\muR^2 = \muGPD^2 = \muDA^2 = \Q^2$. 
The evolution of GPDs and DAs with factorization scale is discussed 
below 
in Sec. \ref{sec:evol}.

In (\ref{eq:Hconvolution}) and (\ref{eq:HVconvolution}) the superscript $a$
denotes the flavor of the parton off which hard scattering occurs,
and the sum over active flavors is implied.
Different flavor compositions of individual mesons make it, in principle, possible to separate
the contribution of individual flavors to GPDs, which is
one of the main reasons for studying the DVMP process. 
Flavor decompositions of CFFs 
have been systematically organized
in \cite{Kumericki:2007sa} (Sec. 3.1),
while TFFs specifically for various mesons
are systematized in \cite{Mueller:2013caa} (Sec. 2.2).
In this work we are interested in neutral vector mesons.
Taking into account the mixing of flavor singlet 
quark and gluon GPDs under evolution,
it is convenient to decompose DVCS CFFs and 
\dvvlpO TFFs into a
flavor nonsinglet (NS) and singlet (S) part,
where
the singlet part is  given as a sum of
gluon (G) and flavor singlet quark ($\Sigma$) contributions:
\begin{subequations}
\begin{eqnarray}
\mathcal{H} 
&=& 
Q^2_{\text{NS}}
\,
\mathcal{H}^{\text{NS}}
+
Q^2_{\text{S}}
\,
\mathcal{H}^{\text{S}}
 = 
Q^2_{\text{NS}}
\,
\mathcal{H}^{\text{NS}}
+
Q^2_{\text{S}}
\,
\left(
\mathcal{H}^{\Sigma}
+
\mathcal{H}^{\text{G}}
\right) \,,
 \\[0.3cm]
\mathcal{H}_{\PVLO} 
&=& 
\mathcal{H}_{\PVLO}^{\text{NS}}
+
Q_{\PVLO}
\,
\mathcal{H}_{\PVLO}^{\text{S}}
 = 
\mathcal{H}_{\PVLO}^{\text{NS}}
+
Q_{\PVLO}
\,
\left(
\mathcal{H}_{\PVLO}^{\Sigma}
+
\mathcal{H}_{\PVLO}^{\text{G}}
\right)
\, .
\end{eqnarray}
\label{eq:HNSSdecom}
\end{subequations}
The factors
$Q^2_{\text{NS}} = 1/9$ $(1/6)$
and
$Q^2_{\text{S}} = 2/9$ $(5/18)$
depend on the number of active flavors $n_{\rm f} = 3$ $(4)$,
while
$Q_{\PVLO}$ depends on the meson type.
In this work we use 
$n_{\rm f} = 4$,
and
$Q_{\rho^0}=1/\sqrt{2}$.
For small $\xb$, following Regge theory and by analogy with standard PDFs,
the contribution of valence quarks is expected to be suppressed by $\sim\!\sqrt{\xb}$,
so we neglect it and
assume that all of the amplitude comes from sea quarks and
gluons. 
We further assume that the quark sea is flavor symmetric and given
by the singlet combination
\begin{equation}
\label{eq:defHxSigma}
H^{\Psea}(x, \ldots) \approx
H^{\Sigma}(x, \ldots) \equiv \sum_{\Pq=\Pqu, \Pqd, \Pqs, \ldots} 
H^{\Pq(+)}
\equiv \sum_{\Pq=\Pqu, \Pqd, \Pqs, \ldots} 
H^{\Pq}(x, \ldots) - H^{\Pq}(-x, \ldots) 
\;,
\end{equation}
and there is no nonsinglet contribution in our phenomenological analysis.

The symmetry properties of the subprocess hard-scattering amplitudes 
and the meson DA project GPDs of well defined signature $\sigma$, 
as defined in \cite{Mueller:2013caa}, Eqs. (3.1), (3.9) and (3.10)%
\footnote{
For quark GPDs, $\sigma=+1(-1)$ denotes antisymmetric (symmetric) 
behavior under $x \to -x$, and the reverse holds for gluon GPDs. 
GPDs with well-defined symmetry under $x \to -x$ exchange 
are distinguished by superscripts $(\pm)$. 
These superscripts indicate the charge parity of two partons, 
which, when multiplied by the signature, gives the intrinsic parity. 
Consequently, $H^{q, G(+)}$ and $\widetilde{H}^{q, G(+)}$ 
contribute to DVCS, while $H^{q, G(+)}$ and $\widetilde{H}^{q(-)}$ 
contribute to DV production of neutral vector and pseudoscalar mesons, 
respectively. The same holds for $E$ and $\widetilde{E}$.
}
.
In \cite{Mueller:2013caa} the corresponding subprocess hard-scattering
amplitudes were further simplified using the symmetry properties of 
quark (antisymmetric) and gluon (symmetric) GPDs and the vector meson DA (symmetric),
and the use of GPDs of definite signature is understood. 
In \cite{Duplancic:2016bge} an analogous treatment of DV production of pseudoscalar mesons is described.
Consequently, 
in the following
the superscript $a$
in (\ref{eq:Hconvolution}) and (\ref{eq:HVconvolution}), 
apart from information on flavor content, carries the information
on signature. 

The convolutions $\stackrel{x}{\otimes}$ and $\stackrel{v}{\otimes}$ 
in equations (\ref{eq:Hconvolution}) and (\ref{eq:HVconvolution}) 
entail integrations over the longitudinal momentum fractions of the partons, 
as expressed by:
\begin{subequations}\label{eq:convolution}
\begin{align}
    T(x, \xi, \ldots) \stackrel{x}{\otimes} H(x, \xi, \ldots)
    &\equiv  \int_{-1}^{1} \frac{\d x}{2\xi} \: T(x, \xi, \ldots) H(x, \xi, \ldots) \,, 
     \\
T(v, \ldots) 
\stackrel{v}{\otimes} 
    \varphi(v, \ldots) 
    &\equiv  \int_{0}^{1} \d v \: 
T(v, \ldots)  
\varphi(v, \ldots) 
\,. 
\end{align}
\end{subequations}
However, in this work, we apply a transformation from the longitudinal momentum space 
$x$ into the space of conformal moments $j$ ($j$-space) to all hard-scattering amplitudes, GPDs, and DAs.
For singlet quark and gluon GPDs such a transformation
is given by a convolution with Gegenbauer polynomials $C_{j}^{\nu}$
\begin{subequations}
\label{eq:defHj} 
\begin{align}
    H_{j}^{\PSigma}(\xi, t, \mu_{\rm GPD}^2)& = 
\frac{\Gamma(3/2)\Gamma(1+j)}{2^{j} \Gamma(j+3/2)}\;
\frac{1}{2}\;
\int_{-1}^{1}\d x\: \xi^j \,
C_{j}^{3/2}(x/\xi) \, H^{\PSigma}(x, \xi, t,\mu_{\rm GPD}^2) \;,
\label{eq:defHjSigma} \\
    H_{j}^{\PG}(\xi, t, \mu_{\rm GPD}^2)& = 
\frac{\Gamma(3/2)\Gamma(1+j)}{2^{j} \Gamma(j+3/2)}\;
\frac{1}{2}\;
\int_{-1}^{1}\d x\: 
\frac{3}{j} \;
\xi^{j-1} \,
C_{j-1}^{5/2}(x/\xi) \, H^{\PG}(x, \xi, t, \mu_{\rm GPD}^2) \,,
\label{eq:defHjg}
\end{align}
\end{subequations}
for integer $j$,
where the normalization coefficients are expressed via Euler's gamma function. 
For (the quark) meson DA the corresponding transformation is for an integer $k$ also
given by a convolution with Gegenbauer polynomials
\begin{equation}
    \varphi_{k}(\mu_{\rm DA}^2) = \frac{2(2k+3)}{3(k+1)(k+2)} \int_{0}^{1} \d v\: 
    C_{k}^{3/2}(2v-1) \varphi(v, \mu_{\rm DA}^2) \,.
    \label{eq:defphik}
\end{equation}
The normalizations of GPD and DA conformal moments differ.
The former ensures that in forward kinematics
($t=0$ and $\xi=0$) conformal moments of GPDs
coincide with corresponding PDF Mellin moments,
while the latter is fixed by
\begin{equation}
    \varphi(v, \mu_{\rm DA}^2) = 
6 v (1-v) \sum_{k}^\infty
    \varphi_{k}(\mu_{\rm DA}^2) \;
    C_{k}^{3/2}(2v-1) \,.
\end{equation}
The symmetry properties of GPDs and the DA, 
whether symmetric or antisymmetric under 
$x \to -x$ and $v \to 1 - v$, are transferred 
to even and odd parities of conformal moments.

As singlet quarks $\PSigma$ and gluons $\PG$ 
mix under evolution with scale, 
all relations for DVCS and \dvvlpO are effectively
matrices in $(\PSigma, \PG)$ space. Thus the transformation of the hard-scattering amplitude
into the space of conformal moments can be written as
\begin{subequations}
\label{eq:ConMomCjjk}
\begin{multline}
\label{eq:ConMomC}
 \mbox{\boldmath $T$}_{j}
(\ldots) 
= \frac{
2^{j+1}\Gamma(j+5/2)}{\Gamma(3/2) \Gamma(j+4)}
 \frac{1}{2} \int_{-1}^1\! dx\;\mbox{\boldmath $T$}
(x, \xi=1, \ldots) \\
\times \left(
\begin{array}{cc}
 (j+3)\big[1-x^2\big] C_j^{3/2}(x) & 0 \\
0 & 3\big[1-x^2\big]^2 C_{j-1}^{5/2}(x)
\end{array}
\right)\,,
\end{multline}
and in the case of \dvvlpO additionally
\begin{equation}
 \mbox{\boldmath $T$}_{jk}
(\ldots) 
    = 6 \int_{0}^{1} \d v\: 
    v (1-v) \, C_{k}^{3/2}(2v-1) 
 \mbox{\boldmath $T$}_{j} (v, \ldots)
 \,.
    \label{eq:ConMomCjk}
\end{equation}
\end{subequations}

Applying these transformations, the convolution for $\mathcal{H}^{\text{S}}$ becomes an infinite
divergent sum
\begin{equation}
\label{def:conmomsum}
\mathcal{H}^{\text{S}}\big(\xb,t,\Q^2\big) =
 2 \sum_{j=1\atop {\rm odd}}^\infty \xi^{-j-1}
 \mbox{\boldmath $T$}^{\rm DVCS}_{j} \left(\alphas(\muR),
        \frac{\Q^2}{\muR^2}, \frac{\Q^2}{\muGPD^2} \right)
\; \mbox{\boldmath $H$}_{j}(\xi,t,\muGPD^2) 
\;,
\end{equation}
and similarly for $\mathcal{H}_{\PVLO}^{\text{S}}$, with additional summation over the
meson DA conformal moments $\Sigma_{k=0, {\rm even}}^\infty$.
In accordance with the CPaW formalism, 
we perform the resummation of the sum \req{def:conmomsum} 
through Mellin-Barnes integration in the complex plane. 
This yields the final expression for DVCS CFF $\mathcal{H}^{\text{S}}$
\begin{equation}
\label{eq:MBH} \mathcal{H}^{\text{S}}(\xb,t,\Q^2)
 = \frac{1}{2 i}\int_{c-i \infty}^{c+ i \infty}
dj\; \xi^{-j-1} \left[i + \tan
\left(\frac{\pi j}{2}\right)\right] 
 \mbox{\boldmath $T$}^{\rm DVCS}_{j} 
\left(\alphas(\muR), \frac{\Q^2}{\muR^2}, \frac{\Q^2}{\muGPD^2} \right)
\; \mbox{\boldmath $H$}_{j}(\xi,t,\muGPD^2) \,,
\end{equation}
and, completely analogously, for DVMP TFF $\mathcal{H}_{\PVLO}^{\text{S}}$
\begin{eqnarray}
\mathcal{H}_{\PVLO}^{\text{S}}(\xb,t,\Q^2)
    &=& \frac{C_F f_{\PVLO}}{N_c \Q}
 \frac{1}{2 i}\int_{c-i \infty}^{c+ i \infty}
dj\; \xi^{-j-1} \left[i + \tan
\left(\frac{\pi j}{2}\right)\right] 
\nonumber \\ & &
\times
\left[
\sum_{k} 
\mbox{\boldmath $T$}^{\dvvlpO}_{jk} 
\left(\alphas(\muR), \frac{\Q^2}{\muR^2}, \frac{\Q^2}{\muGPD^2}, 
\frac{\Q^2}{\muDA^2} \right)
\varphi_{\PVLO,k}
\left(\muDA^2 \right)
\right] \,
\mbox{\boldmath $H$}_{j}(\xi, t,\muGPD^2)\,,
\nonumber \\
    \label{eq:MBHV} 
\end{eqnarray}
where $\mbox{\boldmath $T$}$ and $\mbox{\boldmath $H$}$ are  analytically
continued in the complex $j$ plane.
The intersection $c$ of the integration contour with the real axis should be to the right
of the Regge poles and the poles of anomalous dimensions and hard-scattering amplitudes,
and to the left of the poles of the tangent function.
In this paper we use the value $c=0.35$.
The analogous expressions without gluon contributions are valid for
$\mathcal{H}^{\text{NS}}$ and
$\mathcal{H}_{\PVLO}^{\text{NS}}$
from \req{eq:HNSSdecom}. 
As mentioned earlier, these contributions are negligible for the kinematics
of interest and have not been considered in our numerical analysis.

We have elucidated here the transition from
the usual momentum fraction representation 
to the conformal moment representation and 
the CPaW formalism,
where
the form factors
\req{eq:Hconvolution} and \req{eq:HVconvolution} 
are expressed in the form
\cite{Kumericki:2007sa,Mueller:2013caa}
\begin{subequations}
\begin{eqnarray}
    \mathcal{H}(\xb, t, \Q^2) 
& = &  
H^a_j(\xi,t, \muGPD^2)
\:\stackrel{j}{\otimes}\: 
{}^aT^{\rm DVCS}_j\left(\alphas(\muR),
        \frac{\Q^2}{\muR^2}, \frac{\Q^2}{\muGPD^2} \right) \,,
                             \label{eq:Hcpwe} 
\end{eqnarray}

\begin{eqnarray}
\lefteqn{\mathcal{H}_{\PVL}(\xb, t, \Q^2) }
\nonumber \\
&=& \frac{C_{\rm F} f_{\PVL}}{N_c \Q}
\:
H^a_j(\xi, t, \muGPD^2)  
\:\stackrel{j}{\otimes}\: 
{}^aT^{\dvvlp}_{jk}
\left(\alphas(\muR),
\frac{\Q^2}{\muR^2}, \frac{\Q^2}{\muGPD^2}, 
\frac{\Q^2}{\muDA^2} 
\right)
\:\stackrel{k}{\otimes}\:
 \varphi_{\PVL,k}(\muDA^2) 
\:
\, ,
\nonumber \\
              \label{eq:HVcpwe}
\end{eqnarray}
\end{subequations}
where
\begin{subequations}
\begin{align}
    {}^aT_j(\ldots) \stackrel{j}{\otimes} H^a_j(\ldots)
    &\equiv  
 \frac{1}{2 i}\int_{c-i \infty}^{c+ i \infty}
dj\; \xi^{-j-1} 
\left[ 
i \pm 
\left\{ {\tan \atop \cot} \right\} \left(\frac{\pi j}{2}\right)
\right] 
\: {}^aT_j (\ldots) H^a_j(\ldots) 
 \label{eq:cpwe} 
\\
 & \hspace*{2cm}  \mbox{for } \sigma(H^a) =\left\{ {+1 \atop -1} \right\}
\,, 
 \nonumber \\
T_k(\ldots) 
\stackrel{k}{\otimes} 
    \varphi_k(\ldots) 
    &\equiv  
\sum_{k=0\atop {\rm even}}^\infty
T_k(\ldots)  
\varphi_k(\ldots) 
\,. 
 \label{eq:cpweb} 
\end{align}
\end{subequations}
Here one employs $\tan$ or $\cot$ in accordance with signature $\sigma$.
The summation over $k$, i.e., DA conformal moments,
can, in principle, also be replaced by a Mellin-Barnes integral
when the analytical form is known.
We will provide a summary of the relevant conformal moments 
(\ref{eq:ConMomCjjk}) 
for DVCS and \dvvlp subprocess hard-scattering amplitudes in the following section.

\subsection{Hard-scattering amplitudes}
\label{sec:hscoef}

\subsubsection{DVCS}

\begin{subequations}
\label{eq:CjDVCS}
The DVCS hard-scattering amplitude 
contributing 
to the singlet intrinsic parity even (vector) CFF
\req{eq:MBH}
and determined to NLO order in $\msbar$ scheme 
reads in the conformal moment space
\cite{Kumericki:2006xx,Kumericki:2007sa}
\begin{equation}
\label{eq:TDVCS}
\mbox{\boldmath $T$}^{\textnormal{DVCS}}_{j}
\left(\alphas(\muR),
\frac{\Q^2}{\muR^2}, 
\frac{\Q^2}{\muGPD^2} 
\right)
= \frac{2^{j+1} \Gamma(j+5/2)}{\Gamma(3/2)\Gamma(j+3)}
\left[{\mbox{\boldmath $C$}_{j}^{(0)}}  +
    \frac{\alphas(\muR)}{2\pi} \mbox{\boldmath $C$}_j^{ (1)}
\left( \Q^2/\muGPD^2 \right)
          +  {\cal O}(\alphas^2) \right],
\end{equation}
with the LO part simply
\begin{equation}
    {\mbox{\boldmath $C$}_{j}^{(0)}} = \left({}^{\PSigma}C_{j}^{(0)},{}^{\PG}C_{j}^{(0)}\right) = (1, 0) \;,
    \label{eq:CDVCS0}
\end{equation}
and the NLO part
\begin{eqnarray}
    {}^{\PSigma}C_j^{(1)}
\left( \Q^2/\muGPD^2 \right)
&\!\!\!=\!\!\!&  C_{\rm F} \left[ 2
    S^2_{1}(1 + j)- \frac{9}{2}  + \frac{5-4S_{1}(j+1)}{2(j + 1)_{2}} 
+ \frac{1}{[(j+1)_{2}]^2}\right] - \frac{{}^{\PSigma\PSigma}\gamma_j^{(0)}}{2}
\ln\frac{{\Q}^2}{\muGPD^2} \, ,
\qquad
\\
    {}^{\PG}C_j^{(1)}
\left( \Q^2/\muGPD^2 \right)
&\!\!\!=\!\!\!& -n_{\textup{f}}\frac{(4 + 3j + j^2)
\left[S_{1}(j)+S_{1}(j+2)\right]  +2 + 3j + j^2}{
(1 + j)_{3} }
- \frac{{}^{\PSigma \PG}\gamma_j^{(0)}}{2}
\ln\frac{{\Q}^2}{\muGPD^2} 
\, . \qquad
\label{eq:C1DVCS}
\end{eqnarray}
\end{subequations}
The parameter $n_{\textup{f}}$ is the number of active quark flavors, $(z)_{n}$ is
the Pochhammer symbol
\begin{equation}
    (z)_{n} = \frac{\Gamma(z+n)}{\Gamma(z)} = \prod_{k=0}^{n-1} (z+k) \,,
    \label{eq:poch}
\end{equation}
$S_1(z)$ is the harmonic number
\begin{equation}
    S_1(z) = \frac{\d}{\d z} \ln \Gamma(z+1) + \gamma_{\rm E} \;,
    \label{eq:defS1}
\end{equation}
with Euler-Mascheroni constant $\gamma_{\rm E} = \num{0.57721566}$,
while ${}^{\Sigma\Sigma}\gamma_j^{(0)}$ and ${}^{\Sigma \PG}\gamma_j^{(0)}$ are
elements of the LO anomalous dimensions matrix, given below 
in (\ref{eq:gammas}).

The hard-scattering amplitude contributing to $\mathcal{H}^{\text{NS}}$, 
which we do not include in the numerical evaluation, is, to this order, 
identical to ${}^\Sigma T_j^\dvcs$.
The same subprocess hard-scattering amplitudes contribute to the CFF $\mathcal{E}$, 
which is neglected in the current phenomenological analysis.
Similarly, in this work we do not consider 
the parity-odd (axial-vector) hard-scattering amplitudes 
contributing to
$\widetilde{\mathcal H}$ and $\widetilde{\mathcal E}$ TFFs. 
These amplitudes are listed in 
\cite{Kumericki:2007sa} (Section 4.2).

\subsubsection{DVMP}

Unlike DVCS, 
for DVMP the LO is already proportional to $\alphas$ 
and in the singlet sector of interest, gluons contribute at LO
(Fig. \ref{fig:DVMPDVCS}).
The perturbative expansion
of the \dvvlpO
hard-scattering amplitude 
contributing to the singlet intrinsic parity even (vector) 
TFF
\req{eq:MBHV} 
takes the form \cite{Mueller:2013caa,Duplancic:2016bge}
\begin{subequations}
\begin{eqnarray}
\label{eq:TDVMP}
\lefteqn{\mbox{\boldmath $T$}^{\dvvlpO}_{jk}
\left(\alphas(\muR),
\frac{\Q^2}{\muR^2}, 
\frac{\Q^2}{\muGPD^2}, 
\frac{\Q^2}{\muDA^2} 
\right)} 
\nonumber \\[0.2cm] &=&
 3 \: \frac{2^{j+1} \Gamma(j+5/2)}{\Gamma(3/2)\Gamma(j+3)}
\left[{\alphas
(\muR) 
\mbox{\boldmath $T$}_{j k}^{(0)}}  +
    \frac{\alphas^2(\muR)}{2\pi} 
\mbox{\boldmath $T$}_{j k}^{ (1)}
\left(\Q^2/\muR^2, \Q^2/\muGPD^2, \Q^2/\muDA^2 \right)
          +  {\cal O}(\alphas^3) \right],
\quad
\end{eqnarray}
where factor 3 is a consequence of the normalization of the meson DA, and the
LO term reads 
\begin{equation}
    {\mbox{\boldmath $T$}_{j k}^{(0)}} =
    \left({}^{\PSigma} T_{j k}^{(0)}, {}^{\PG} T_{j k}^{(0)}\right) =
    \left(\frac{1}{n_{\textup{f}}}, 
    \:\frac{2}{C_{\textup{F}}(j+3)}\right) \;.
    \label{eq:TDVMP0}
\end{equation}
At the NLO order, the so-called pure singlet contribution ${}^{\textnormal{pS}}T^{(1)}$
appears for the first time
\begin{eqnarray}
\lefteqn{ {\mbox{\boldmath $T$}_{jk}^{(1)}} 
\left( 
\ldots
\right)
}
\nonumber \\[0.2cm]
&=& \left(\frac{1}{n_{\textup{f}}} {}^{\Pq}T_{j k}^{(1)} 
\left( 
\frac{\Q^2}{\muR^2}, 
\frac{\Q^2}{\muGPD^2}, 
\frac{\Q^2}{\muDA^2} 
\right)
+ 
    {}^{\textnormal{pS}} T_{j k}^{(1)}
\left( 
\frac{\Q^2}{\muGPD^2} 
\right)
,\: \frac{2}{C_{\textup{F}}(j+3)}{}^{\PG}T_{j k}^{(1)}
\left( 
\frac{\Q^2}{\muR^2}, 
\frac{\Q^2}{\muGPD^2}, 
\frac{\Q^2}{\muDA^2} 
\right)
\right) 
\;.
\nonumber \\
    \label{eq:TDVMP1}
\end{eqnarray}
Note that ${}^{\Pq}T_{j k}$ 
corresponds to the nonsinglet part that contributes
to the, in our numerical analysis neglected, TFF $\mathcal{H}_{\PVLO}^{\text{NS}}$
from \req{eq:HNSSdecom}.
It is convenient to separate amplitudes $T_{j k}^{(1)}$ according to contributions of
individual classes of Feynman diagrams,
i.e., to make a color decomposition
\begin{align}
    {}^{\Pq}T_{j k}^{(1)} 
\left( 
\ldots
\right)
&= C_{\rm F} \, c_{jk}^{(1, {\rm F})} 
\left( 
\frac{\Q^2}{\muGPD^2}, 
\frac{\Q^2}{\muDA^2} 
\right)
+ \beta_0 \, c_{jk}^{(1, \beta)}  
\left( 
\frac{\Q^2}{\muR^2} 
\right)
    + C_{\rm G}\,  c_{jk}^{(1, {\rm G})} \,, 
\label{eq:Tqjk}
\\[0.2cm]
{}^{\textnormal{pS}}T_{j k}^{(1)}
\left( 
\ldots
\right)
 &= {}^\textnormal{pS}c_{jk}^{(1)} 
\left( 
\Q^2/\muGPD^2 
\right)
\,, \\[0.2cm]
    {}^{\PG}T_{j k}^{(1)} 
\left( 
\ldots
\right)
&= C_{\rm F} \,{^{\PG} c}_{jk}^{(1, {\rm F})}  
\left( 
\frac{\Q^2}{\muGPD^2}, 
\frac{\Q^2}{\muDA^2} 
\right)
+ C_{\rm A} \, {}^{\PG} c_{jk}^{(1, {\rm A})}
\left( 
\frac{\Q^2}{\muGPD^2} 
\right)
+ \frac{\beta_0}{2} \, 
 \ln\frac{\mu_{\textnormal{GPD}}^2}{\muR^2} 
\,,
\end{align}
\end{subequations}
where 
$C_{\rm F} = 4/3$, 
$C_{\rm A} = 3$, 
$C_{\rm G} =C_{\rm F}-C_{\rm A}/2= -1/6$, 
and 
\begin{equation}
    \beta_0 = \frac{2}{3} n_{\textup{f}} - \frac{11}{3}C_{\rm A}
\, .
\label{eq:beta0}
\end{equation}
Conformal moments are given in \cite{Mueller:2013caa}:
the nonsinglet moments
$c_{jk}^{(1, \textup{F})}$, $c_{jk}^{(1, \beta)}$
and $c_{jk}^{(1, \textup{G})}$ 
in Eq. (4.44),
the pure singlet moments
${}^\textnormal{pS}c_{jk}^{(1)}$ in Eq. (4.48), 
and the gluon moments
${}^{\PG}c_{jk}^{(1, \textup{F})}$ and ${}^{\PG}c_{jk}^{(1, \textup{A})}$ 
in Eq. (4.53).
However, as was stated in \cite{Duplancic:2016bge}, the factorization
of the gluon contribution employed in \cite{Mueller:2013caa} 
(as well as in preceding work \cite{Ivanov:2004zv,Ivanov:2004vd}),
was a nonstandard one%
, i.e., differed from the one used in the case of DIS and DVCS. 
Hence, corrections were required 
for both the pure singlet and gluon contributions%
\footnote{
In the erratum \cite{Ivanov:2004zv, Ivanov:2004vd}, 
only the quark contributions were corrected. 
The mixing of quark and gluon contributions under evolution 
leads to corrections in both contributions: 
the correction of the NLO quark contribution depends 
on the LO gluon contribution, and vice versa. 
Thus, in the erratum \cite{Ivanov:2004zv}, 
the correction to the gluon contribution was overlooked. 
In contrast, due to the vanishing LO quark contribution, 
the erratum \cite{Ivanov:2004vd} for the $J/\Psi$ production is correct, 
as confirmed also by \cite{Jones:2015nna}.
}
, 
as elucidated in \cite{Duplancic:2016bge} 
(specifically in Eq. (20) and below).
For convenience we list the corrected expressions 
for the pure singlet moment
\begin{subequations}
\begin{eqnarray}
\label{eq:cpS}
{}^\textnormal{pS}{c}^{(1)}_{jk}
\left( 
\ldots
\right)
&\!\!\!=\!\!\!&
\left[-\ln\frac{\Q^2}{\mu_{\textnormal{GPD}}^2} - 1 + 2 S_ 1(j+1) + 
2S_ 1(k+1)-1\right] 
\frac{{}^{\PG \PSigma}\gamma_j^{(0)}}{C_{\text{F}}(j+3)}
\nonumber
\\ &&\!\!\!\!\!\!
- \left[\frac{1}{2}+\frac{1}{(j+1)_2}+\frac{1}{(k+1)_2}\right]\frac{2}{(j+1)_2} + 
{}^\textnormal{pS}\Delta c^{(1)}_{jk}
\, ,
\end{eqnarray}
and the gluon conformal moment
\footnote{
In addition to the omission noted in \cite{Duplancic:2016bge}, 
during the work on this article, 
in one of the terms in ${}^{\PG}c_{jk}^{(1, {\rm F})}$
in (\protect\ref{eq:cjkGF})
a typo in the sign was identified
and corrected.}
\begin{eqnarray}
    \label{eq:cjkGF}
{}^{\PG}{c}^{(1,\text{F})}_{jk} 
\left( 
\ldots
\right)
&\!\!\!=\!\!\!&
\left[
  -\ln\frac{\Q^2}{\mu_{\textnormal{DA}}^2}+S_ 1(j+1)+S_1(k+1) 
  - \frac{3}{4}-\frac{1}{2(k+1)_2}-\frac{1}{(j+1)_2}
\right] \frac{{}^{\PSigma \PSigma}\gamma_k^{(0)}}{2 C_{\text{F}}}
 \\ &&\!\!\!\!\!
 +\left[-\ln\frac{\Q^2}{\mu_{\textnormal{GPD}}^2} + 1 + 3 S_1(j+1) 
 - \frac{1}{2}  + \frac{2 S_1(j+1)-1}{(k+1)_2} -\frac{1}{(j+1)_2}\right]
 \frac{j+3}{2}
 \frac{{}^{\PSigma \PG}\gamma_j^{(0)}/n_{\text f}}{2}
\nonumber \\ &&\!\!\!\!\!
-\left[35 - \left[(k+1)_2+2\right] \Delta S_2\Big(\frac{k+1}{2}\Big) + 
\frac{4}{[(k+1)_2]^2}\right]\frac{1}{8}
\nonumber \\ &&\!\!\!\!\!
+\left[\frac{\left[(k+1)_2+2\right] S_ 1(j+1)}{(k+1)_2} +1 \right] 
\frac{1}{(j+1)_2}+ {}^{\PG}\Delta{c}^{(1,\text{F})}_{jk}
\,.
\nonumber
\end{eqnarray}
\end{subequations}
These replace ${}^\textnormal{pS}{c}^{(1)}_{jk}$ and 
${}^{\PG}{c}^{(1,\text{F})}_{jk}$ in Eqs. (4.48) and (4.53b)
in \cite{Mueller:2013caa}.
Here
$\Delta S_{2}$ is defined  as in \cite{Mueller:2013caa}, Eq. (4.13),
while
${}^\textnormal{pS}\Delta c^{(1)}_{jk}$
and
${}^{\PG}\Delta{c}^{(1,\text{F})}_{jk}$
are given in \cite{Mueller:2013caa}, 
Eqs. (4.48b) and (4.53f), respectively.
The anomalous dimensions
 ${}^{\PG \PSigma}\gamma_j^{(0)}$,
 ${}^{\PSigma \PG}\gamma_j^{(0)}$
and ${}^{\PSigma \PSigma}\gamma_k^{(0)}$
are to be found in the next section.

Finally, the same subprocess hard-scattering amplitudes contribute to the 
TFF $\mathcal{E}_{V_L^0}$,
which is neglected in the current phenomenological analysis.
The hard-scattering amplitudes for deeply virtual production
of pseudoscalar mesons 
associated 
with $\widetilde{H}$ and $\widetilde{E}$ GPDs,
are listed in \cite{Duplancic:2016bge}.

\subsection{Evolution of GPDs and DAs}
\label{sec:evol}

In the context of the longitudinal momentum fraction space 
(often referred to as $x$-space), 
GPD evolution at LO has been implemented in computer code 
for a long time \cite{Vinnikov:2006xw}. 
However, this implementation has not yet been widely utilized for global fits. 
There is optimism that a new approach, as developed in \cite{Bertone:2022frx}, 
when combined with the PARTONS framework \cite{Berthou:2015oaw}, 
may lead to advancements in this area. 
Additionally, there exists an implementation of NLO evolution 
in $x$-space \cite{Freund:2001bf}, 
but this code has proven to be challenging to use effectively.
One of the main advantages of working in the space of conformal moments
($j$-space) 
is the simplicity and numerical stability it offers for GPD evolution,
which enables us to work at full NLO level.
In conformal moment representation the evolution operator 
is diagonal at LO, and conformal moments evolve autonomously. 
Closed analytical expressions are available
for both diagonal and non-diagonal terms appearing
at NLO and beyond (for detailed account see \cite{Kumericki:2007sa}). 
The CPaW framework thus offers the potential 
for direct extension to the NNLO level. 
In the case of DVCS, preliminary analysis has 
been conducted in a specialized conformal scheme, 
leveraging DIS NNLO results 
\cite{Mueller:2005nz, Kumericki:2006xx, Kumericki:2007sa}. 
Additionally, there is already a significant number of components 
available in the $\msbar$ scheme 
\cite{Braun:2020yib, Braun:2021grd, Ji:2023xzk, Braun:2017cih, Braun:2022bpn}.

\subsubsection{Evolution operators}
The operator $\bEop (\mu, \muO;\xi)$ 
governing the perturbative evolution 
of the singlet intrinsic parity even (vector) GPDs of interest in this work,
can be defined as
\begin{equation}
\left( \begin{array}{c}
        {H_j^{\PSigma}(\xi,t,\mu^2)} \\[0.2cm]
        {H_j^{\PG}(\xi,t,\mu^2)} \\
      \end{array}
\right)  =
\bEop_{jl}(\mu,\muO;\xi)
\left( \begin{array}{c}
        {H_l^{\PSigma}(\xi,t,\mu_{0}^2)} \\[0.2cm]
        {H_l^{\PG}(\xi,t,\mu_{0}^2)} \\
      \end{array}
  \right) \,,
\label{eq:evS}
\end{equation}
where at NLO accuracy and in the $\msbar$ scheme, the operator is non-diagonal
in both the two-dimensional flavor space $(\PSigma, \PG)$ and in
the infinite-dimensional 
space of conformal moments.
We can write the perturbative expansion of this operator in the form
\begin{eqnarray}
\label{Exp--EvoOpeSin} 
\lefteqn{ \bEop_{jl}(\mu,\muO; \xi)
} \nonumber \\
&=&
 \sum_{a,b=\pm}\left[
    \delta_{ab}\, {\mbox{\boldmath $P$}}^{a}_j \delta_{jl} +
      \frac{\alpha_s(\mu)}{2\pi}
     \left( {\mbox{\boldmath ${\cal A}$}}^{(1)ab}_j(\mu,\muO)
     \delta_{jl} +
      {\mbox{\boldmath ${\cal B}$}}^{(1)ab}_{jl}(\mu,\muO)
     \, \xi^{j-l}\right)
+O(\alphas^2) \right]\left[ \frac{\alphas(\mu)}{\alphas(\muO)}
\right]^{-\frac{{\lambda}^{b}_l}{\beta_0}}
\, . \nonumber\\
\end{eqnarray}
Here, the summation is performed over the eigenstates $a, b \in \{+, -\}$ of the LO
evolution operator in $(\PSigma, \PG)$ space, where the projectors
${\mbox{\boldmath $P$}}^{a}_j$ onto those states are
\begin{eqnarray}
\label{Def-ProP}
{\mbox{\boldmath $P$}}^{\pm}_j = \frac{\pm 1}{{
\lambda}^{+}_j-{\lambda}^{-}_j} \left(\mbox{\boldmath
$\gamma$}_j^{(0)}-{\lambda}^{\mp}_j \mbox{\boldmath
$1$}\right)\,.
\end{eqnarray}
Here $\mbox{\boldmath $\gamma$}_j^{(0)}$ is a $2 \times 2$ matrix
of anomalous dimensions 
whose elements 
for the intrinsic parity even (vector) case
read \cite{Gross:1974cs,Georgi:1974wnj}
\begin{subequations}
\label{eq:gammas}
\begin{eqnarray}
\label{eq:gam0}
{}^{\PSigma \PSigma}\gamma_{j}^{(0)\,} &\!\!\!=&\!\!\!
- C_{\rm F} \left( 3 + \frac{2}{(
j + 1 )( j + 2 )} - 4 S_{1}(j + 1) \right) \,, \label{eq:gam0QQ}\\
{}^{\PSigma \PG}\gamma_{j}^{(0)\,} &\!\!\!=&\!\!\! -2n_{\textup{f}} 
\frac{4 + 3\,j + j^2 }{( j + 1 )( j + 2 )( j + 3)}\,,
\label{eq:gam0QG} \\
{}^{\PG \PSigma}\gamma_{j}^{(0)\,} &\!\!\!=&\!\!\!
-2C_{\rm F}\frac{4 + 3\,j + j^2 }{j( j + 1 )( j + 2 )}\,,
\label{eq:gam0GQ}\\
{}^{\PG \PG}\gamma_{j}^{(0)\,}
   &\!\!\!=&\!\!\! - C_{\rm A} \left(-\frac{4}{( j + 1 )( j + 2
)}+\frac{12}{j( j + 3)} - 4S_1( j + 1 )  \right)+ \beta_0\,,
\label{eq:gam0GG}
\end{eqnarray}
\end{subequations}
while
\begin{eqnarray}
\label{Def-EigVal}
{\lambda}^{\pm}_j
 =\frac{1}{2}
\left({}^{\PSigma\PSigma}\gamma_j^{(0)} 
+ {}^{\PG \PG}
\gamma^{(0)}_j \mp
 \left({}^{\PSigma\PSigma} \gamma^{(0)}_j 
- {}^{\PG\PG} \gamma^{(0)}_j \right)
\sqrt{1 + \frac{4\, 
{}^{\PSigma\PG} \gamma^{(0)}_j 
\;
{}^{\PG\PSigma}\gamma^{(0)}_j}{\left(
{}^{\PSigma\PSigma} \gamma^{(0)}_j 
- {}^{\PG\PG} \gamma^{(0)}_j\right)^2 }}
\right)\,,
\end{eqnarray}
are its eigenvalues.
The diagonal part of the NLO evolution operator is given as
\begin{equation}
\label{Def-A1} {\mbox{\boldmath ${\cal A}$}}_j^{(1)ab}
= R^{ab}_{jj}(\mu,\muO)\, {\mbox{\boldmath
$P$}}^{a}_j \left[ \frac{\beta_1}{2\beta_0}  \mbox{\boldmath
$\gamma$}_j^{(0)} -\mbox{\boldmath $\gamma$}_j^{(1)} \right]
{\mbox{\boldmath $P$}}^{b}_j \,,
\end{equation}
with
\begin{equation}
{R}^{ab}_{jl}(\mu,\muO)
= \frac{1}{ \beta_0+{\lambda}^{a}_j-{\lambda}^{b}_l}\left[
1- \left(\frac{\alphas(\muO)}{\alphas(\mu)}\right)^{\frac{
\beta_0+{\lambda}^{a}_j-{\lambda}^{b}_l}{\beta_0}} \right]
\, ,
\end{equation}
$\beta_1$ specified below (\ref{eq:asrge}), and where the NLO anomalous dimensions $\mbox{\boldmath $\gamma$}_j^{(1)}$
can be read from \cite{Floratos:1981hs, Belitsky:1998uk} with the convention
change specified in \cite{Kumericki:2007sa} (Eq. (134)).
The non-diagonal part of the NLO evolution operator, which is the part that leads to
mixing of conformal moments, can be written in the form
\begin{eqnarray}
{\mbox{\boldmath ${\cal B}$}}_{jl}^{(1)ab}(\mu,\muO)= -
R^{ab}_{jl}(\mu,\muO) \; \left(
\lambda^{a}_j-\lambda^{b}_l \right) \left[ \left(
\beta_0-\lambda^{b}_l \right) \mbox{\boldmath $P$}^{a}_j
\;\mbox{\boldmath $d$}_{jl} \mbox{\boldmath $P$}^{b}_l +
\mbox{\boldmath $P$}^{a}_j \; \mbox{\boldmath $g$}_{jl}
\mbox{\boldmath $P$}^{b}_l \right] \,,
\end{eqnarray}
where the matrices $\mbox{\boldmath $d$}_{jl}$ and $\mbox{\boldmath $g$}_{jl}$
can be read off \cite{Belitsky:1998uk} with the same convention
change as applied to the diagonal anomalous dimension matrices.
The expressions for the flavor nonsinglet sector
can be derived from the ones provided above by reducing the
matrix-valued quantities to scalar values associated with
quark contributions, i.e., $\PSigma \PSigma$.
Thus, one gets the nonsinglet evolution operator 
$\Eop (\mu,\muO;\xi)$
by employing the replacements outlined in \cite{Kumericki:2007sa}, Eq. (144).
We note that to NLO accuracy considered here
\begin{equation}
{}^{\rm NS}\gamma_{j}^{(0)}
=
{}^{\PSigma \PSigma}\gamma_{j}^{(0)}
\, ,
\quad
{}^{\rm NS}\gamma_{j}^{(1)}
=
{}^{\PSigma \PSigma}\gamma_{j}^{(1)}
\, .
\end{equation}

The evolution of the DA is governed by the same evolution operators as for GPDs, 
with skewness $\xi=1$ (ERBL evolution \cite{Efremov:1979qk,Lepage:1980fj}),
while properly accounting for the parity and charge 
quantum numbers. 
Therefore, for the vector mesons under consideration in this work, 
the evolution is governed by the
nonsinglet intrinsic parity even evolution operator $\Eop$.
\begin{eqnarray}
\label{eq:ev-DA}
\varphi_k(\mu^2) = \Eop_{km}(\mu,\muO) \, \varphi_m(\mu^2_0)
\end{eqnarray}
with
\begin{equation}
\gamma_k
=
{}^{\rm NS}\gamma_k
\, ,
\quad
 \Eop_{km}(\mu,\muO) 
=
 \Eop_{km}(\mu,\muO;1) 
\, .
\end{equation}

The evolution of the intrinsic parity odd (axial) GPDs 
and of the pseudoscalar meson DA
follow analogously, and the corresponding anomalous dimensions
are specified in \cite{Kumericki:2007sa}.

\subsubsection{Evolution implementation}

Finally, the evolution of GPDs and consequent convolution with the
hard-scattering amplitude
can be equivalently interpreted as the 
evolution of the hard-scattering amplitude%
\footnote{
We note that the term 'evolution' in this context 
does not involve the resummation of $\Q^2/\mu^2$ logarithms --
just $\mu^2/\muO^2$ terms are resummed.
This should be viewed as an evaluation technique 
and the hard-scattering amplitude retains a residual 
dependence on the factorization scale $\mu^2$. 
For a more detailed discussion of factorization dependence, 
see \cite{Melic:2001wt}.}
and a consequent convolution with GPDs
\begin{equation}
    \mbox{\boldmath $T$}_{l}(\mu)  \stackrel{l}{\otimes} \mathbb{E}_{lj}(\mu,\mu_0) 
    \stackrel{j}{\otimes} \mbox{\boldmath $H$}_{j}(\mu_0) \;.
    \label{eq:TEH}
\end{equation}
This is practical for numerical evaluation, where evolved hard-scattering
amplitudes can be stored in computer memory and called up during fitting of GPDs.
Summation over conformal moment $l$ in (\ref{eq:TEH}) 
can also be replaced by Mellin-Barnes integration, 
and after some reshuffling of the terms 
one gets
\begin{align}
    \mbox{\boldmath $T$}_{l}(\ldots\frac{\Q^2}{\mu^2}\ldots) 
\stackrel{l}{\otimes}\: 
\mbox{\boldmath $H$}_{l}(\ldots\mu^2) 
    &=
\Tevol_{j}
\left(
\ldots
\frac{\Q^2}{\mu^2} 
\ldots;
\{
\mu,
\muO
\}
\right)
    \stackrel{j}{\otimes} 
\mbox{\boldmath $H$}_{j}(\ldots\muO^2) 
\end{align}
with
$\stackrel{j}{\otimes}$
defined in \req{eq:cpwe},
and
\begin{eqnarray}
\label{Cal-OffDigEvo}
\Tevol_{j}
\left(
\ldots
\frac{\Q^2}{\mu^2} 
\ldots;
\{
\mu,
\muO
\}
\right)
&\!\!\!=\!\!\!& 
\mbox{\boldmath $T$}_{j}(\ldots\frac{{\Q}^2}{\mu^2}\ldots)\, 
\bEop_{jj}(\mu,\muO;1)
\\
&& -\frac{1}{4i} \int_{c'-i\infty}^{c'+i\infty}\! dl\,
\cot\left(\frac{\pi l}{2}\right) \mbox{\boldmath
$T$}_{j+l+2}(\ldots\frac{{\Q}^2}{\mu^2}\ldots)\, 
\bEop_{j+l+2,j}(\mu,\muO;1)
\,,
\nonumber
\end{eqnarray}
with $-2 < c' < 0$.
In this work we use $c'=-0.25$.
The form specified in \req{Cal-OffDigEvo} is applicable 
to both intrinsic parity even and odd GPDs, regardless of their signatures. 
This is because the integral over $l$ stems from the generic sum 
$\Sigma_{l=0 , {\rm even}}^\infty$.
Naturally, the hard-scattering amplitudes differ,
and the operators include the corresponding vector
or axial-vector anomalous dimensions, as elaborated
in \cite{Kumericki:2007sa}.
The nonsinglet expression is analogous.

The distribution amplitude is at some input scale $\muO$
often represented in terms of the finite number of 
conformal moments \req{eq:defphik}.
The evolved DA \req{eq:ev-DA}
is then to LO also presented by a finite sum.
However, NLO evolution introduces mixing between
conformal moments and an infinite sum emerges.
In practical applications, this infinite sum is 'truncated' 
assuming the suppression of higher conformal moments.
When the analytical form of DA conformal moments
is known, Mellin-Barnes integration
can be employed, resulting in an expression similar to
\req{Cal-OffDigEvo}.
For DVMP, it is advantageous to 'precalculate' 
both DA and GPD evolutions with the hard-scattering amplitude
\begin{eqnarray}
\lefteqn{
\Tevol_{jk}
\left(
\ldots
\frac{\Q^2}{\muGPD^2}, 
\frac{\Q^2}{\muDA^2} 
;
\{
\muGPD,
\mu_{0}
\},
\{
\muDA,
\mu_{0}'
\}
\right)
\qquad
}
\nonumber \\
&=& 
\left[
\mbox{\boldmath $T$}_{lm}
\left(
\ldots,
\frac{\Q^2}{\muGPD^2}, 
\frac{\Q^2}{\muDA^2} 
\right)
\stackrel{m}{\ootimes}
 \Eop_{mk}(\muDA, \mu_{0}')
\right]
\stackrel{l}{\ootimes}
\bEop_{lj}
(\muGPD, \muO; \xi)
\, ,
\end{eqnarray}
where 
$\stackrel{l}{\ootimes}$
corresponds to
\req{Cal-OffDigEvo}, 
while
$\stackrel{m}{\ootimes}$
can be a summation over $m$ or an analogue of
\req{Cal-OffDigEvo}. 
Often one takes 
$\muO=\mu_{0}'$.

Finally, the CFF and TFF expressions
suitable for practical use 
are given by
\begin{subequations}
\label{eq:HHVcpweEV} 
\begin{eqnarray}
    \mathcal{H}^{\text S}(\xb, t, \Q^2) 
& = &  
\Tevol_j^\dvcs
\left(
\ldots
\frac{\Q^2}{\muGPD^2} 
\ldots;
\{
\muGPD,
\muO
\}
\right)
    \stackrel{j}{\otimes} 
\mbox{\boldmath $H$}_{j}(\ldots\muO^2) 
\, ,
\label{eq:HcpweEV} 
\end{eqnarray}
\begin{eqnarray}
\lefteqn{\mathcal{H}_{\PVL}^{\text S}(\xb, t, \Q^2) }
\nonumber \\
&=& \frac{C_{\rm F} f_{\PVL}}{N_c \Q}
\left[
\Tevol_{jk}^{\dvvlpO}
\left(
\ldots
\frac{\Q^2}{\muGPD^2}, 
\frac{\Q^2}{\muDA^2} 
;
\{
\muGPD,
\muO
\},
\{
\muDA,
\mu_{0}'
\}
\right)
\:\stackrel{k}{\otimes}\:
 \varphi_{\PVL,k}(\mu_{0}'^{2}) 
\right]
    \stackrel{j}{\otimes} 
\mbox{\boldmath $H$}_{j}(\ldots\muO^2) 
\:
\, ,
\nonumber \\
\label{eq:HVcpweEV}
\end{eqnarray}
\end{subequations}
where
$\stackrel{j}{\otimes}$
is given by
\req{eq:cpwe},
while 
$\stackrel{k}{\otimes}$
is a summation 
\req{eq:cpweb} 
or
\req{eq:cpwe} as well.
Moreover, expressions (\ref{eq:HHVcpweEV}) allow us to quantify
the influence of quarks and gluons
present at initial scale $\muO$.
Note that due to the fact that
$\bEop$ 
\req{Exp--EvoOpeSin} 
is nondiagonal in the flavor space
$(\PSigma, \PG)$, 
the components
${}^{\PSigma} \tevol$ and  ${}^{\PG} \tevol$ 
represent the mixture of
${}^{\PSigma} T$ and  ${}^{\PG} T$. 
Consequently,
in contrast to the gluon
contribution at scale $\mu$
obtained using \req{eq:TDVCS},
the contribution
${}^{\PG} \tevol^\dvcs_j
\stackrel{j}{\otimes}
{}^{\PG} H_j(\ldots\muO^2)$
is different than zero even at LO.

As is well-known in renormalization group theory, the dependence on the
evolution scale is primarily manifested through the scale-dependent behavior of
the strong coupling constant. In numerical evaluations, we utilize the
equation
\begin{equation}
\frac{d a_s}{d \ln \mu^2} = \beta_0 a_s^2 + \beta_1 a_s^3 \,,
\label{eq:asrge}
\end{equation}
where $a_s = {\alphas}/(4\pi)$, $\beta_0$ is given in (\ref{eq:beta0}),
and $\beta_1 = 10C_{\rm A}n_{\rm f}/3 + 2 C_{\rm F}n_{\rm f} - 34 C_{\rm A}^2 / 3$.
At LO we employ the analytical solution considering only the first term in (\ref{eq:asrge}),
while at NLO we utilize the numerical fourth-order Runge-Kutta integration. 
At the initial scale $\mu^2 = \qty{2.5}{\GeV^2}$, we take the values 0.0606 (LO) 
or 0.0518 (NLO) for $\alphas(\mu)/(2\pi)$, and set $n_{\rm f} = 4$, 
consistent with the kinematical range of interest here.

\subsection{Understanding contributions: simple lessons and insights}
\label{sec:lessons}

When examining the expressions presented in this section within the realm of phenomenology, 
it is crucial to note the following key points:
\begin{itemize}
     \item 
The gluon component of the NLO hard-scattering DVCS amplitude in (\ref{eq:CjDVCS}) 
is negative. Consequently, gluonic contributions at NLO strongly suppress
the quark contributions, as later illustrated in Fig. \ref{fig:QvsGDVCS}. 
The most dominant contribution in (\ref{eq:C1DVCS}) arises from the term:
\begin{equation}
            -n_{\textup{f}} \frac{S_{1}(j+2)}{(1+j)_{3}} \,,
     \label{eq:hsdomterm}
\end{equation}
         which in $x$-space corresponds to the most singular part of the amplitude
         proportional to
\begin{equation}
            \frac{\ln^2 (x-\xi)}{x-\xi} \,.
\end{equation}
The DIS structure function $F_1$ 
corresponds to the imaginary part 
of ${\cal H}$ in the forward limit%
\footnote{This observation was exploited in \cite{Kumericki:2007sa} 
where NNLO DIS Wilson coefficients contributing to $F_1$ 
were utilized to access NNLO DVCS in the special conformal scheme.},
and it's worth noting that precisely \req{eq:hsdomterm} does not contribute 
to its gluon Wilson coefficient 
(\cite{Kumericki:2007sa}, Eq. (105)).
     \item 
In DVMP, gluons appear already at LO \req{eq:TDVMP0},
and owing to their high density 
in the relevant kinematic regime of small $\xb$, which is already established 
from the study of the DIS process, they are responsible for the dominant part 
of the DVMP amplitude. This is evident in Fig. \ref{fig:QvsGDVMP}.
     \item 
The $j=0$ pole, a well-known feature present in the LO gluon anomalous dimensions 
${}^{\PG \PSigma}\gamma_{j}^{(0),}$ 
and ${}^{\PG \PG}\gamma_{j}^{(0),}$ \req{eq:gammas}, 
plays a crucial role in driving strong gluon evolution at small $\xb$. 
This pole represents the $j$-space signature of the interplay 
between the Bjorken $\Q^2 \to \infty$ 
and the high-energy limit $1/\xb \to \infty$. 
The impact of this gluon evolution is evident 
in Figs. \ref{fig:QvsGDIS} and \ref{fig:QvsGDVCS}, 
where the DIS structure function and DVCS CFF 
exhibit an increase with $\Q^2$. 
Additionally, Fig. \ref{fig:QvsGDVMP} illustrates 
the flat behavior of the DVMP TFF in the relevant energy region. 
This flat behavior, as a consequence, facilitates the correct 
scaling behavior demonstrated in Fig. \ref{fig:Q2scaling}.
\item
The $j=0$ pole is not only present in the intrinsic parity 
even evolution operator $\bEop$ \req{Exp--EvoOpeSin} 
but also surfaces in the NLO DVMP contributions 
for the pure singlet moment ${}^\textnormal{pS}c_{jk}^{(1)}$ \req{eq:cpS} 
and the gluon contribution ${}^{\PG}c_{jk}^{(1, \textup{A})}$ 
(\cite{Mueller:2013caa}, Eq. (4.53a)). 
These contributions are parametrized using ${}^{\PG \PSigma}\gamma_{j}^{(0),}$ 
and ${}^{\PG \PG}\gamma_{j}^{(0),}$, 
multiplying the factorization logarithm $\ln \Q^2/\muGPD^2$ 
and constant terms. 
Similar contributions are expected for DVCS at NNLO.
\end{itemize}

\section{Model for GPDs and DA}
\label{sec:gpdmodel}

To complete our theoretical framework, it is essential to specify 
a parameterization of GPDs that facilitates fitting to experimental measurements. 
We adopt a model wherein conformal moments of GPDs, as defined in (\ref{eq:defHj}), 
are characterized by an expansion in $t$-channel SO(3) partial waves 
\cite{Mueller:2005ed,Kumericki:2007sa,Kumericki:2009uq,Muller:2014wxa}. 
The foundational concepts for this approach were laid out in 
\cite{Mueller:2005ed,Kumericki:2007sa} 
and subsequently refined and applied, particularly in the context of small-$\xb$ phenomenology, 
in \cite{Kumericki:2009uq}%
\footnote{The model for GPD $H$ conformal moments for small $\xb$ was utilized in \cite{Kumericki:2009uq}, 
alongside the separate determination of Compton form factors in the valence region. It's worth noting that the KM-model nomenclature is used interchangeably for both results, even though they represent distinct approaches to DVCS phenomenology.}.
In \cite{Muller:2014wxa}, the fundamental steps connecting this $j$-space 
modeling with more conventional $x$-space GPDs were outlined. 
The former approach offers several advantages, including a clear connection 
with Mellin moments and lattice calculations, while the latter is more intuitively clear. 
It's crucial to emphasize that most of the hard-exclusive phenomenology 
and GPD modeling has traditionally been carried out in $x$-space,
see recent examples in \cite{Moutarde:2018kwr,Kriesten:2021sqc}.
Apart from the previously listed applications of the CPaW formalism to 
DIS, DVCS and DVMP at NLO and beyond, 
recently the conformal moment representation has been employed 
at LO \cite{Guo:2022upw,Guo:2023ahv}, aiming to establish a parallel global analysis 
program encompassing PDFs, form factors, and DVCS measurements, 
with a foreseeable extension to lattice data.

We take generic expression \cite{Kumericki:2009uq}
\begin{equation}
    H^{a}_{j}(\xi, t) = \sum_{J = 0 \atop{\rm even} }^{j+1}
     \xi^{j+1-J} \; \hat{d}^{a}_J(\xi) \; H^{a}_{j,J}(t) \;,
    \quad a \in \{\Psea=\PSigma, \PG\} \;,
\label{eq:HPWE}
\end{equation}
where $J$ is the angular momentum in the $t$ channel, and
$\hat{d}^{a}_J(\xi)$ are the corresponding Wigner functions.
As in \cite{Kumericki:2009uq}, we work in the approximation 
$\hat{d}^{a}_{J}(\xi)\approx 1$, which is good enough for small-$\xb$
kinematics.  In accordance with the fact that the forward limit of GPDs is
equal to ordinary PDFs, e.g. for quark $\Pq$,
\begin{equation}
    H^{\Pq}(x, 0, 0) = \theta(x) q^{\Pq}(x) - \theta(-x) \bar{q}^{\Pq}(-x)\,,
    \label{eq:fwd}
\end{equation}
the amplitude $H_{j, J=j+1}(t)$ of the leading partial wave is for $t=0$
equal to the Mellin moment of the corresponding PDF
\begin{equation}
    H^{\Pq}_{j, j+1}(0) = q^{\Pq}_j \equiv  \int_{0}^{1} dx\; x^j q^{\Pq}(x) \;.
\label{eq:def-qj}
\end{equation}
In principle, the values $q^{a}(x)$ could be taken from the results of one of the
collaborations specialized in the extraction of PDFs from experimental
measurements. However, state-of-the-art PDF extraction usually employs
variable flavor number procedures with sophisticated
matching at heavy-quark production thresholds. Thus, naive
usage of numerical values for $q^{a}(x)$ from those studies in our
simplified fixed-flavor-number framework would lead to
inconsistencies. Therefore, as in \cite{Kumericki:2007sa,Kumericki:2009uq}, we
determine the Mellin moments of PDFs ourselves, by fitting a simple ansatz
\begin{equation}
    q^{a}_j = N_{a} \frac{B(1-\alpha^{a}_{0}+j, \beta^{a}+1)}{B(2-\alpha^{a}_{0},\beta^{a}+1)}\;,
\label{eq:qj}
\end{equation}
to DIS measurements, where the counting rules determine $\beta^{\Psea} = 8$ and $\beta^{\PG} = 6$.
This ansatz corresponds to a standard simple parameterization in $x$-space
\begin{equation}
    q^{a}(x) = \frac{N_{a}}{B(2-\alpha^{a}_{0},\beta^{a}+1)} x^{-\alpha_{0}^{a}}
    (1-x)^{\beta^{a}}\;.
\label{eq:qx}
\end{equation}
The normalization in (\ref{eq:qj}) and (\ref{eq:qx}) is chosen so that $N_{a}$
corresponds to the average longitudinal momentum fraction of parton $a$.
These parameters are constrained by the summation rule
\begin{equation}
    N_{\Psea} + N_{\rm val} + N_{\PG} = 1 \,,
    \label{eq:sumrule}
\end{equation}
and where, in accordance with results of DIS studies, we take $N_{\rm val} = 0.4$.
Thus, we use only three parameters to fit small-$\xb$ DIS measurements:
\begin{equation}
 \{ N_{\Psea},\, \alpha_{0}^{\Psea},\, \alpha_{0}^{\PG} \} \,.
    \label{eq:DISpars}
\end{equation}
We complete the model of the leading partial wave by adding the dependence on $t$,
and that, firstly, by completing the full Regge trajectory
\begin{equation}
    \alpha_{0}^a \to \alpha^a(t) = \alpha^{a}_{0} + \alpha'^{a} t\,, 
    \label{eq:regge}
\end{equation}
and, secondly, by adding the residual dependence on $t$ via the dipole
impact factor, so that the complete expression for the leading partial wave is
\begin{equation}
    H^{a}_{j, j+1}(t) \equiv q^{a}_{j}(t) =
    q^{a}_j \frac{1+j-\alpha^{a}_0}{1+j-\alpha^{a}_0 - \alpha'^{a} t}
    \left(1 - \frac{t}{m_{a}^2}\right)^{-2} \;,
\label{eq:qjt}
\end{equation}
with $q^{a}_j$ given in (\ref{eq:qj}).
It turns out that the data we work with cannot distinguish the differences in $t$- and
$j$-dependencies of individual partial waves, so for subleading waves we
assume that they are simply proportional to the leading wave (\ref{eq:qjt}),
\begin{equation}
    H^{a}_{j, J}(t) \propto q^{a}_j(t)\,,
    \label{eq:HjJpropto}
\end{equation}
and at the end we take into consideration
only two subleading waves, so our complete model is given by
\cite{Kumericki:2007sa,Kumericki:2009uq}
\begin{equation}
    H^{a}_j(\xi, t) = \big(1 + s^{a}_{2}\, \xi^2 + s^{a}_{4}\, \xi^4 \big) q^{a}_{j}(t),
\label{eq:lowx-KM-model}
\end{equation}
where normalizations of the subleading partial waves $s^{a}_{2}$ and $s^{a}_{4}$
are additional parameters of the model.
Thus, the total set of parameters, in addition to the three parameters from
(\ref{eq:DISpars}) is given by:
\begin{equation}
    \{\alpha'^{\Psea},\, \alpha'^{\PG},\,
        m_{\Psea}^2,\, m_{\PG}^2,\, 
        s_{2}^{\Psea},\, s_{2}^{\PG},\,
    s_{4}^{\Psea},\, s_{4}^{\PG}  \}
\label{eq:pars}
.\end{equation}
These additional eight parameters are determined by fitting to measured DVCS and DVMP data.

We also mention that the practically identical GPD model was also used in the first multichannel DVCS + DVMP fit in 
\cite{Lautenschlager:2013uya}. The main difference is that the
subleading partial waves were not taken as proportional to the leading one, but,
for example, the Regge parameters $\alpha_0$ and $\alpha'$ were different for each
partial wave. In addition, the overall normalizations of the measured DVMP
cross sections were considered as fitting parameters.
A larger number of parameters makes the model in \cite{Lautenschlager:2013uya} 
more flexible, but the model used in this paper proved itself flexible enough for our needs.

The distribution amplitude 
$\varphi_{\Prhozero}(u)$ is relatively poorly known and
for this study we decided to stick to its asymptotic form, so
the sum over $k$ in (\ref{eq:MBHV}) has only one term
\begin{equation}
    \varphi_{\PVL,0} = 1\,, \qquad \varphi_{\PVL,k>0} = 0 
\;,
    \label{eq:DAasymptotic}
\end{equation}
and NLO evolution, being tiny for this form, is neglected.
More advanced studies could also treat DA as an unknown function whose
lower Gegenbauer moments could be simultaneously fitted to experimental measurements, or,
at this moment probably more convenient, those lower moments could be
taken from lattice QCD results \cite{Braun:2016wnx}.

\section{Experimental data and L/T separation}
\label{sec:exp}

\begin{figure*}[t]
    \centering
    \includegraphics[width=0.8\textwidth]{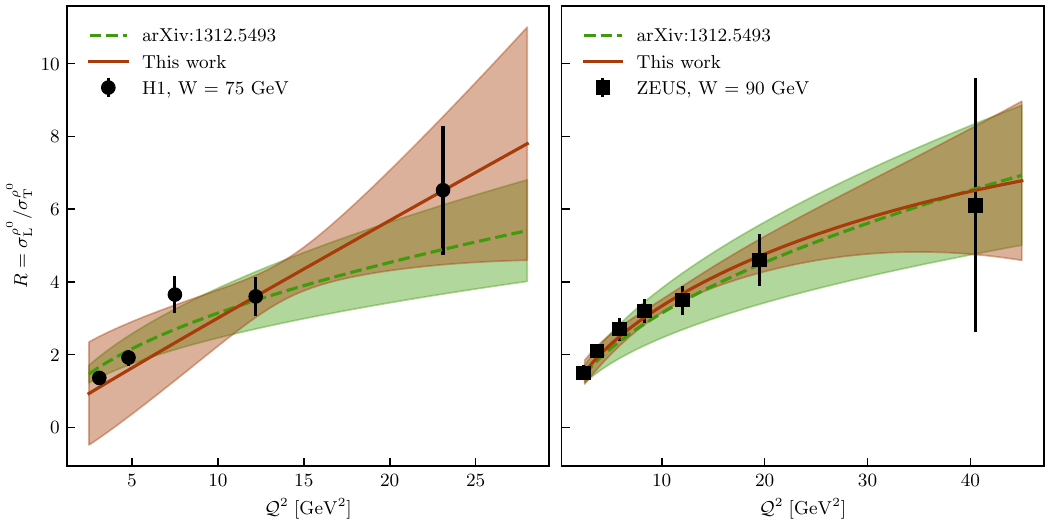}
    \caption{$R(W, \Q^2)$ functions (\protect\ref{eq:Rdef}) for the $\Prhozero$ production obtained by
    separate fits of ansatz (\ref{eq:RWansatz}) to H1 measurements \cite{H1:2009cml} (left),
    and to ZEUS measurements \cite{ZEUS:2007iet} (right) are plotted as the red solid line and
    the corresponding symmetric red uncertainty band.
    The function $R(\Q^2)$ determined by $W$-independent analysis \cite{Lautenschlager:2015mxt}
    using the ansatz (\ref{eq:Ransatz}) is shown for comparison as the green dashed line and the corresponding symmetric green uncertainty band.}
    \label{fig:Rfunction}
\end{figure*}

Since we focus on the high energy regime (small $\xb$)
and large values of virtuality $\Q^2$, the only experimental data that
we use are those of the H1 and ZEUS collaboration realized using the
HERA collider. In particular, we use the results of
\begin{itemize}
  \item H1 measurements of $F_2(\xb, \Q^2)$ DIS structure function from \cite{H1:1996zow}\footnote{This
          data has been superseded by the combined H1 and ZEUS data \cite{H1:2015ubc}. However, these more
          recent measurements are reported in terms of cross-sections that combine all DIS structure functions
          $F_i$, whereas older measurements report $F_2$ so we can use them directly within our framework
          which is limited to describing only the $F_2$ part of the DIS cross section.\label{fn:h115}}
  \item H1 and ZEUS measurements of the DVCS cross section $\sigDVCS \equiv \sigma(\DVCS)$
      from \cite{H1:2005gdw,H1:2009wnw,ZEUS:2003pwh,ZEUS:2008hcd},
  \item H1 and ZEUS measurements of the DVMP cross section for production of the \Prhozero
      meson $\sigma(\DVMPrho)$ from \cite{ZEUS:2007iet,H1:2009cml}.
\end{itemize}
In order to be in the regime where perturbative QCD can be used with confidence
in the leading twist,
for DVCS data, we impose a cut $\Q^2 > \,\qty{5}{\GeV^2}$, and for
DVMP, which is more problematic due to two hadrons in the final state,
we use a data cut $\Q^2 > \,\qty{10}{\GeV^2}$.
For the possible importance of higher-twist contributions see recent results
\cite{Alexeev:2022gkb} at lower $Q^2$.
We could also straightforwardly include measurements of the production of $\Pphi$ mesons.
Neglecting the contribution of valence quarks leads to a simple SU(3) flavor relation
between the cross sections for the production of $\Prhozero$ and $\Pphi$ mesons:
\begin{equation}
    \label{eq:phirho}
    \sigma(\DVMPrho) = \left(\frac{3 f_{\Prho^0}}{\sqrt{2} f_{\Pphi}}\right)^2
    \sigma(\DVMPphi) \;,
\end{equation}
where the meson decay constants are $f_{\Prho^0} = \,\qty{0.209}{\GeV}$ and
$f_{\Pphi} = \,\qty{0.221}{\GeV}$.
However, deviations from this relation are visible in the experimental data in
Fig. \ref{fig:Q2Wscaling}, where the cross sections for $\Pphi$, scaled according to (\ref{eq:phirho}),
are consistently smaller than those for $\Prhozero$. To avoid
tensions that are entirely an artifact of our
simplified model, in this paper we do not use the data for $\Pphi$ (nor for $\Pomega$),
but only those for $\Prhozero$ that have better statistics. Neglected influence
of valence quark contributions results in a small reparametrization of the
sea-parton GPDs. One of the main goals of this study is to establish the
usability of the collinear twist-2 approach to describe the longitudinal
DVMP process\footnote{From this point forward, for the sake of readability, we 
will refer to the process as DVMP, although the technically accurate term is 
DV\HepParticle{\rho}{L}{0}P.},
and that can already be judged on the basis of the $\Prhozero$ meson production alone.

In contrast to the DVCS situation where the dominance of the twist-2 amplitude is a safer
assumption and the measured data can be directly compared with the theoretical prediction,
it is known that the measured DVMP cross section has a large higher-twist contribution from the exchange
of transversely polarized virtual photon. Consequently, for comparison with the twist-2
theory presented in Sec. \ref{sec:pqcd} it is necessary to separate the contributions
of longitudinal and transverse photons (L/T separation) and compare the theory with
measurements of the \emph{longitudinal} cross section
$\sigDVMPL = \sigma(\HepProcess{\Pgamma_{\textup{L}}^*\,\Pproton \to \Prhozero \Pproton})$.
Direct observation of the polarization of the virtual photon is of course not
possible, so L/T separation of the DVMP process is done indirectly, typically by
measuring the polarization of the meson in the final state and using the
assumption of $s$-channel helicity conservation (SCHC), which implies that the polarization
of the photon follows that of the meson,
$\sigDVMPL \approx \sigDVMP = 
\sigma(\HepProcess{\Pgammastar \Pproton \to \Prho_{\textup{L}}^0\,\Pproton})$.
The SCHC assumption can also be experimentally verified by measuring and comparing
individual spin-density matrices elements (SDME), which additionally enables the
improvement of the L/T separation procedure itself.
Using such an approach, both HERA collaborations determined the ratio
\begin{equation}
    \label{eq:Rdef}
    R \equiv \frac{\sigDVMPL}{\sigDVMPT} \,,
\end{equation}
and studied its dependence on the kinematic variables of the process. For example,
the results in \cite{H1:2009cml} show
\begin{itemize}
     \item a clear dependence of $R$ on $\Q^2$,
     \item lack of apparent dependence of $R$ on $W$ within uncertainties, and
     \item certain indications of dependence of $R$ on $t$, visible for higher values of $\Q^2$.
\end{itemize}
Furthermore, in \cite{H1:2009cml}, the H1 collaboration specifically determined
values for $\sigDVMPL$ and $\sigDVMPT$. Unfortunately,
the values of $\sigDVMPL$ are given in a binned form only in $\Q^2$, and for a quality extraction of GPDs
it is important to know the dependence on other kinematic variables, $W$ and $t$.
Therefore, in this work we mainly rely on measurements of the quantity $R$ and, using it together
with measurements of the total cross section $\sigma^{\Prhozero}$, we determine the values
of $\sigDVMPL$ which we use in our fits. Since $R$ is mostly not given for the
same kinematic points as $\sigma^{\Prhozero}$, measured values for $R$ cannot be used
directly, but we interpolate them by fitting, thus obtaining $R$ as a continuous
function of kinematic variables.

In \cite{Meskauskas:2011aa}, motivated by the fact that only the dependence on $\Q^2$ is unquestionable, an ansatz for the function $R$ is proposed of the form
\begin{equation}
    \label{eq:Ransatz}
    R(\Q^2) = \frac{\Q^2}{\mrho^2}
    \left(1 + a \frac{\Q^2}{\mrho^2}\right)^{-p} \;,
\end{equation}
where $\mrho=\,\qty{0.776}{\GeV}$, and $p$ and $a$ are the fitting parameters.
In \cite{Meskauskas:2011aa,Lautenschlager:2013uya,Lautenschlager:2015mxt}
these parameters were determined by fitting to the H1 and ZEUS values for $R$.
In the present work, we have reexamined the dependence of $R$ on kinematic variables
$W$ and $t$. First, as our interest is focused on data having large $\Q^2$,
we observe that precisely at these values the uncertainty of $R$ is greatest,
and thus it is the least clear whether or not there is a dependence on $W$ (cf.
Fig. 39 in \cite{H1:2009cml}). Also, precisely at higher values of
$\Q^2$ both H1 and ZEUS measurements, which are at different $W$, show more significant
divergence in values for $R$, see Fig. \ref{fig:Rfunction}. Therefore,
for the purposes of this work, we made a new determination of the function $R$, where we
extended the ansatz (\ref{eq:Ransatz}) by a factor that also describes the dependence on $W$
\begin{equation}
    \label{eq:RWansatz}
    R(W, \Q^2) = \frac{\Q^2}{\mrho^2}
    \left(1 + a \frac{\Q^2}{\mrho^2}\right)^{-p}
    \left(1 + b \frac{\Q^2}{W}\right) \;,
\end{equation}
where $b$ is the new parameter. For the numerical efficiency of the fitting procedure,
$W$ is not squared in the denominator, which makes the parameter $b$ dimensional.
Also, because of tension between the measurements,
we determined the function $R$ separately
for H1 data, and for ZEUS data:
\begin{align}
    a& = 3 \pm 29\,,& p& = 1.0 \pm 0.6\,,& b& = -82 \pm 780 \,\unit{\GeV^{-1}}\,,& \text{(H1)}\,; \\
    a& = 2 \pm 4\,,& p& = 0.44 \pm 0.26\,,& b& = 0.3 \pm 1.1 \,\unit{\GeV^{-1}}\,,& \text{(ZEUS)}\,.
\end{align}
Thus obtained functions $R(W, \Q^2)_{\textup{H1}}$ and
$R(W, \Q^2)_{\textup{ZEUS}}$ are used in this
work to determine $\sigDVMPL$ from H1 and ZEUS measurements for $\sigma^{\Prhozero}$, respectively.
They are shown in Fig. \ref{fig:Rfunction}.

The question of dependence of $R$ on the variable $t$ also arises. The
situation is again problematic for large values of $\Q^2$, even more than in
the case of the dependence on $W$. Precisely at higher values of $\Q^2$ a dependence on $t$
begins to appear, and additionally, measurements of the production of light
vector mesons at slightly higher $x_B$
measured by the COMPASS collaboration (cf. Fig. 11b in \cite{COMPASS:2020zre})
unambiguously indicate that $R$ also depends on $t$.
On the other hand, we cannot determine this functional dependence because
measurements of $R(t)$ are available only for smaller values of $\Q^2$,
below the limit $\Q^2 = \,\qty{10}{\GeV^2}$ that we set in this paper.
In this unsettled situation, we decided to take a conservative stance and
we do not use DVMP differential cross section measurements $d\sigDVMP/dt$,
but only measurements of the cross section integrated over $t$.

As discussed above, one of the important features of the DVMP
cross section is its scaling with $\Q^2$. Being able to correctly reproduce the scaling
is an important test of a theoretical model.  The scaling of the total
cross section $\sigma^{\Prho^0}$ was usually considered in the literature, but here we
consider the scaling of its longitudinal component $\sigDVMPL$.
A fit of the function $\sigDVMPL \propto \Q^{-w}$ to the data from the left panel of
Fig. \ref{fig:H1_dvmp} gives
\begin{equation}
    \label{eq:Q2powerW}
    w = 5.1 \pm 0.1  \qquad  \text{(fixed $W$)}\;.
\end{equation}
We will see later that NLO models (unlike LO) have no problem reproducing this value.
However, what is usually considered ``the scaling'' is the scaling with $\Q^2$ for fixed $\xb$. Four
measurements of the H1 collaboration at $\xb = 0.0018$ are shown in Fig. \ref{fig:Q2scaling} and fitting the power function
$\sigDVMPL \propto \Q^{-w}$ to these data gives
\begin{equation}
    \label{eq:Q2powerx}
    w = 3.8 \pm 0.2  \qquad  \text{(fixed $\xb$)}\;.
\end{equation}
Such a considerable deviation of scaling
from the asymptotic value $w=6$ is considered to be difficult to describe
within the collinear perturbative QCD approach. However, we find that our LO and NLO models
are capable of reproducing this effective behavior in the measured data region.

\section{Results and discussion}
\label{sec:results}
\begin{figure}[t]
    \centering{\includegraphics[width=0.5\linewidth]{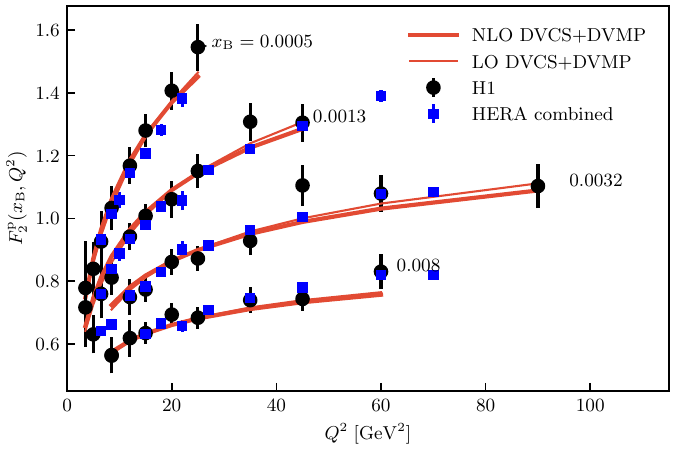}}
    \caption{Description of the DIS structure function $F_2(\xb, Q^2)$ at
        LO (red, thin) and NLO (red, thick) by
     models obtained by a multichannel fit to DIS, DVCS and DVMP data.
     The black points are measurements of the H1 collaboration \cite{H1:1996zow},
     while the blue squares are combined measurements of the H1 and ZEUS collaborations \cite{H1:2015ubc}.}
    \label{fig:H1_dis}
\end{figure}

\begin{figure}[t]
    \centering{\includegraphics[width=0.85\linewidth]{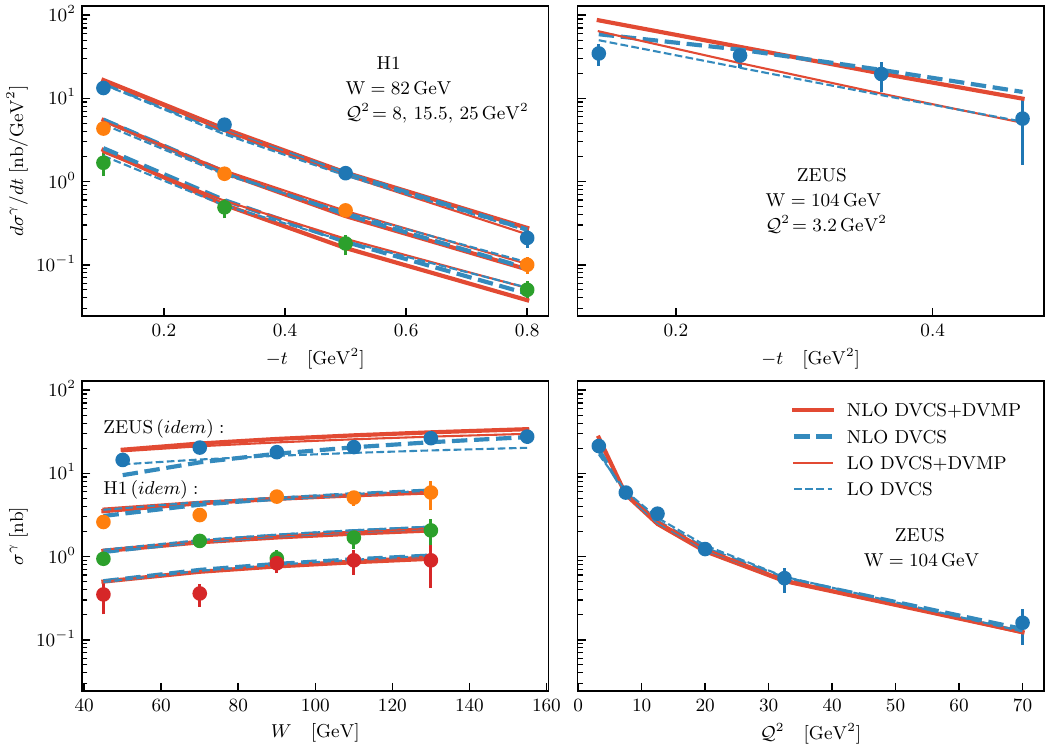}}
    \caption{Description of H1 \cite{H1:2005gdw,H1:2009wnw} and ZEUS 
        \cite{ZEUS:2003pwh,ZEUS:2008hcd} measurements of the dependence of 
        the DVCS cross section on $t$, $W$ and
        $\Q^2$ compared to LO (thin) and NLO (thick) models which are, besides DIS, fitted
        only to these DVCS measurements (blue dashed), and also additionally to
        DVMP measurements (red solid). The three H1 lines on left panels correspond, from top
        down, to $\Q^2 =$ 8, 15.5, and 25\,\unit{GeV$^2$}, respectively.
        }
    \label{fig:H1ZEUS_dvcs}
\end{figure}

\begin{figure}[t]
    \centering{\includegraphics[width=0.8\linewidth]{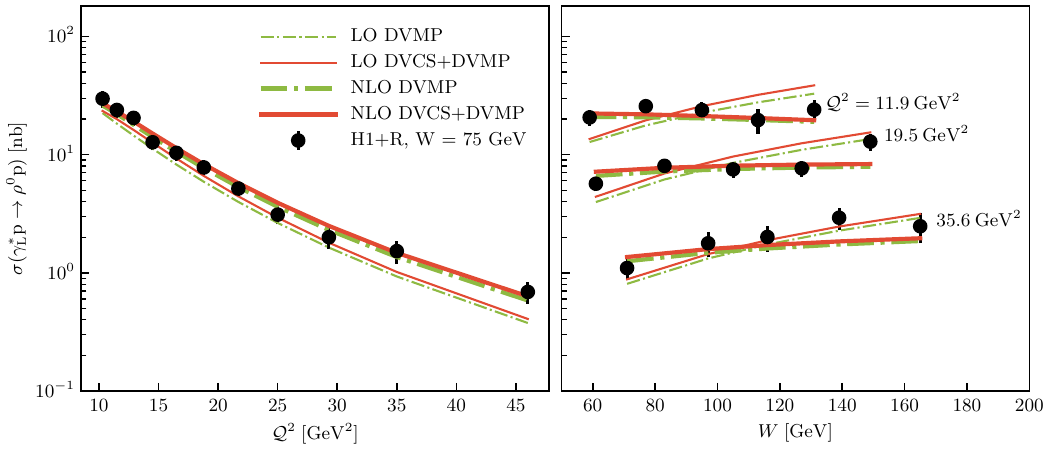}}
    \caption{Description of H1 DVMP measurements \cite{H1:2009cml} by models fitted
     to H1 and ZEUS DVMP data (green, dot-dashed lines) and additionally to DVCS
     data (red, solid lines).  It can be seen that, unlike the NLO models (thick
     lines), the LO models (thin lines) have problems describing the data at larger
     values of $\Q^2$.}
    \label{fig:H1_dvmp}
\end{figure}
\begin{figure}[t]
    \centering{\includegraphics[width=0.8\linewidth]{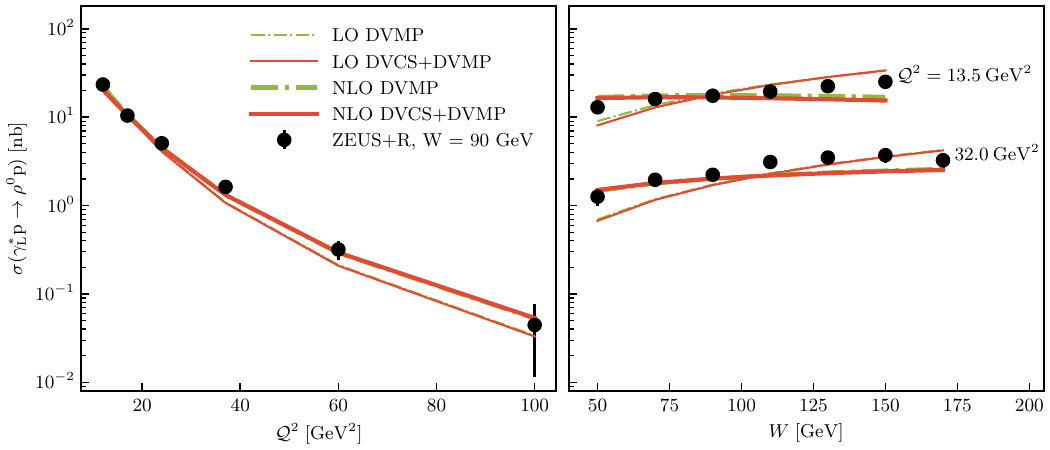}}
    \caption{Description of ZEUS DVMP measurements \cite{ZEUS:2007iet}. Meaning of lines
      is the same as in Fig. \protect\ref{fig:H1_dvmp}.}
    \label{fig:ZEUS_dvmp}
\end{figure}
\begin{figure}[th]
    \centering
    \includegraphics[width=0.55\textwidth]{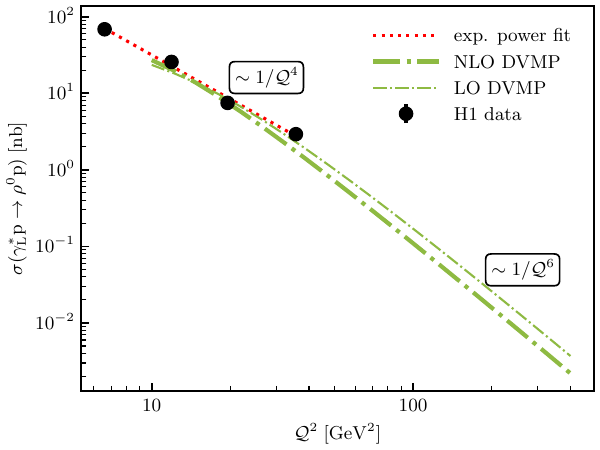}
    \caption{Longitudinal cross sections for production of $\Prhozero$ mesons
        \cite{H1:2009cml}, at fixed $\xb = 0.0018$,  
        together with \model{LO-DVMP} (thin green dot-dashed line)
        and \model{NLO-DVMP} (thick green dot-dashed line) models fitted to DVMP data.
        Both models are able to reasonably reproduce the effective experimental $\Q^4$ scaling in
        data region.}
    \label{fig:Q2scaling}
\end{figure}
\begin{figure}[th]
    \centering
    \includegraphics[width=0.55\textwidth]{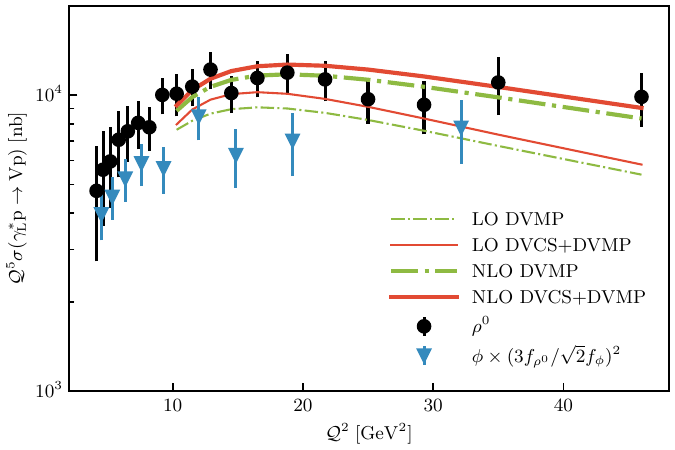}
    \caption{Longitudinal cross sections for production of $\Prhozero$ mesons (black circles)
        \cite{H1:2009cml}, plotted according to the approximate effective experimental
        fixed-$W$ scaling 
        $\sigDVMPL \sim 1/\Q^5$, together with LO (thin lines)
        and NLO (thick lines) models fitted to DVMP only (green dot-dashed) 
        and DVMP+DVCS data (red solid lines).
        One observes that only NLO models are able to reproduce the experimental large $\Q^2$ scaling.
    We also display the SU(3) flavor rescaled $\Pphi$ production data (blue triangles)
    which was not used in this work.}
    \label{fig:Q2Wscaling}
\end{figure}

\begin{figure}[t]
    \centering{\includegraphics[width=\linewidth]{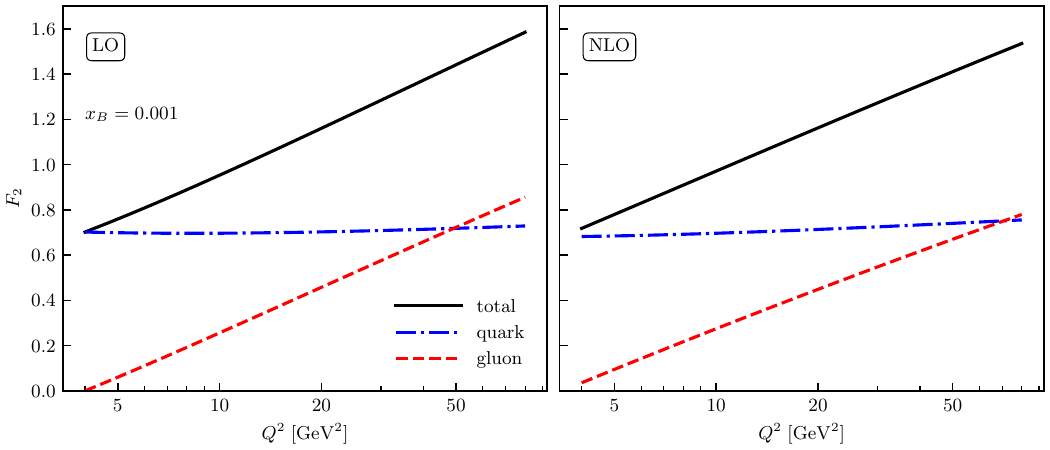}}
    \caption{Contributions of the quark (blue dot-dashed) and gluon (red dashed) PDF to the total
      DIS structure function $F_2$ (black solid) at LO (left) and NLO (right),
      for $\xb = 0.001$.}
    \label{fig:QvsGDIS}
\end{figure}

\begin{figure}[t]
    \centering{\includegraphics[width=\linewidth]{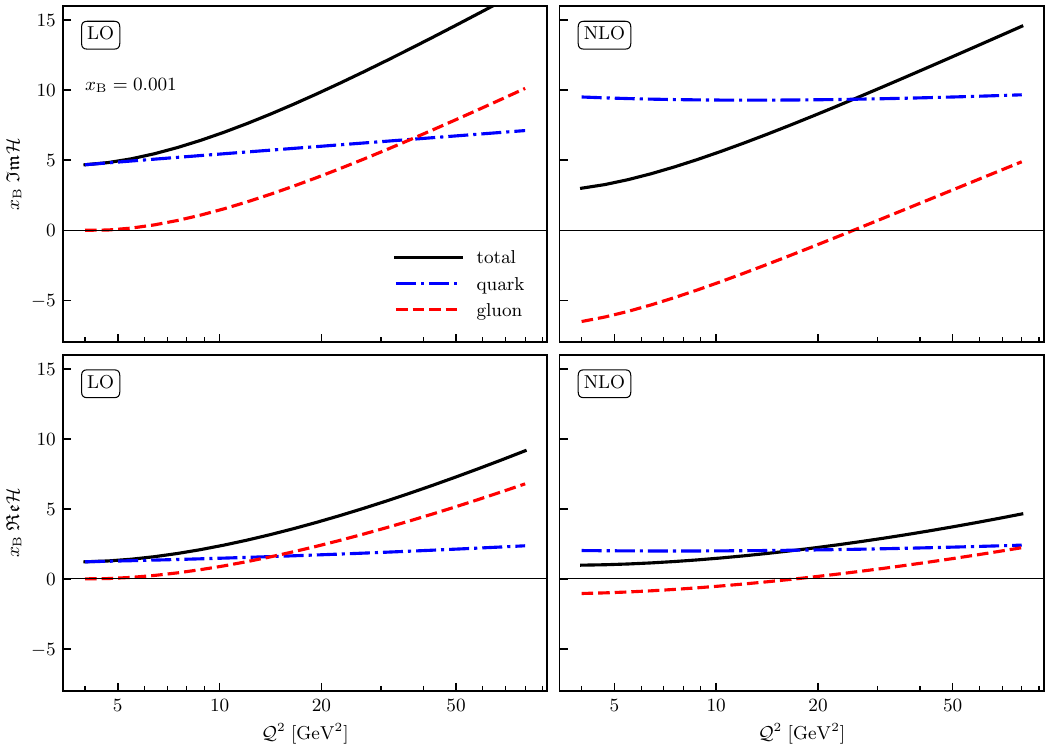}}
    \caption{Contribution of quark (blue dot-dashed) and gluon (red dashed) GPD to the total
        DVCS CFF (black solid) in models \model{LO-DVCS-DVMP} (left) and \model{NLO-DVCS-DVMP}
        (right), in dependence on $\Q^2$ for $\xb = 0.001$ and $t = 0$.
    Imaginary (top) and real (bottom) part of the leading CFF $\mathcal{H}$ are displayed.}
    \label{fig:QvsGDVCS}
\end{figure}

\begin{figure}[t]
    \centering{\includegraphics[width=\linewidth]{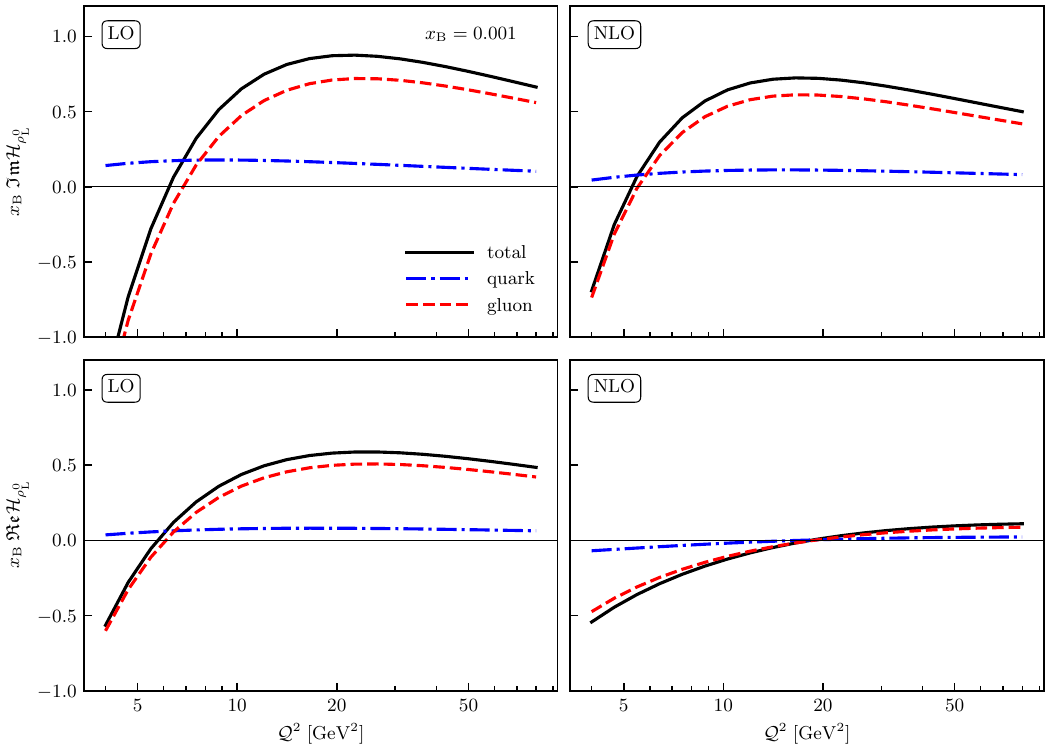}}
    \caption{Contributions of the quark (blue dot-dashed) and gluon (red dashed) GPD to the total
        DVMP TFF (black solid) in models \model{LO-DVCS-DVMP} (left) and \model{NLO-DVCS-DVMP} (right)
      in dependence on $\Q^2$ for $\xb = 0.001$ and $t = 0$.
      Imaginary (top) and real (bottom) part of the leading 
      TFF $\mathcal{H}_{\rho^{0}_{\rm L}}$ are displayed.
      The steep behavior below $\Q^2 = \,\qty{10}{\GeV^2}$ is a reflection of strong
      evolution effects. It lies below our data range and is likely outside
      the scope of validity for our twist-2 collinear approach.
  }
    \label{fig:QvsGDVMP} 
\end{figure}

\begin{figure}[t]
    \centering{\includegraphics[width=\linewidth]{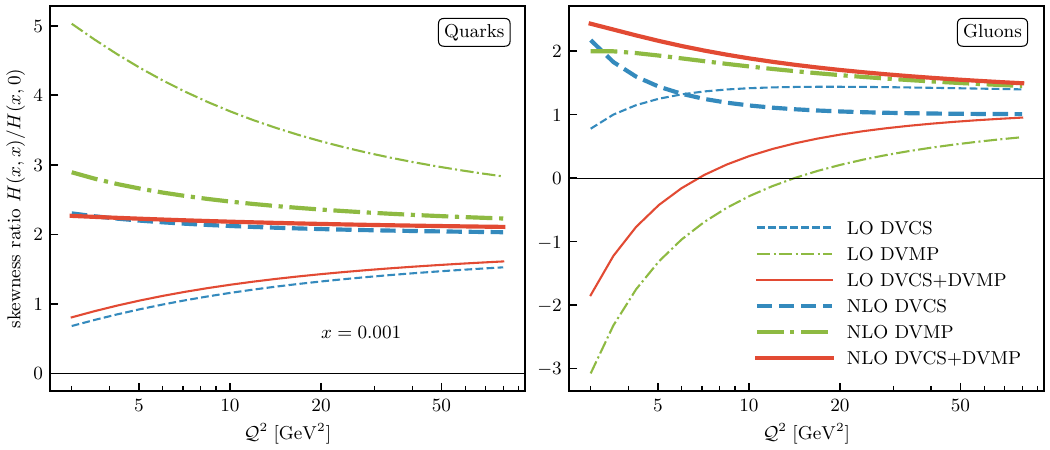}}
    \caption{Skewness ratio 
\protect\req{eq:skewness}
for the quark (left) and gluon (right) GPD $H$,
       for six models from Tab. \protect\ref{tab:models}. 
       Depending on the processes
       used, the LO models (thin lines) show large
       mutual differences of the resulting GPDs. The dependence of NLO GPDs
       (thick lines) on the process is significantly smaller.}
    \label{fig:skewness} 
\end{figure}

\subsection{Fits}

\begin{table} 
\renewcommand{\arraystretch}{1.4}
\small
\caption{\label{tab:models}
     List of models used in this work, pQCD order used, and datasets to which
     the model was fitted. References to used DIS, DVCS and DVMP
     experimental data are given in Tab. \protect\ref{tab:chis}.
}
\centering
\setlength{\tabcolsep}{3pt}
\renewcommand{\arraystretch}{1.2}
\begin{tabular}{ccccc}
\hline\noalign{\smallskip}
model name & order & DIS & DVCS & DVMP  \\
\hline\noalign{\smallskip}
\model{LO-DVCS} & LO & \checkmark & \checkmark &   \\
\model{LO-DVMP} & LO & \checkmark &  & \checkmark  \\
\model{LO-DVCS-DVMP} & LO & \checkmark & \checkmark  & \checkmark  \\
\model{NLO-DVCS} & NLO & \checkmark & \checkmark &   \\
\model{NLO-DVMP} & NLO & \checkmark &  & \checkmark  \\
\model{NLO-DVCS-DVMP} & NLO & \checkmark & \checkmark  & \checkmark  \\
\noalign{\smallskip}\hline
\end{tabular}
\renewcommand{\arraystretch}{1.}
\end{table}

In order to understand the influence of NLO corrections of different processes,
we use six different GPD fits, listed in Tab. \ref{tab:models}.
All of them were obtained by fitting the ansatz described in Sec. \ref{sec:gpdmodel},
at LO or NLO of perturbative QCD. Every model is
first fitted to the DIS $F_2(\xb, \Q^2)$ data of the H1 collaboration \cite{H1:1996zow}. After that,
we fix the parameters (\ref{eq:DISpars}) that affect the form of GPDs in the forward limit, and the
remaining parameters (\ref{eq:pars})
of the three LO and three NLO models are fitted to only DVCS data, only DVMP data and
to the total DVCS+DVMP set, as specified in Tab. \ref{tab:models}.
For the sake of brevity, in the names of all models
we suppressed the DIS label, so, for example, model \model{NLO-DVCS-DVMP}
should have been called \model{NLO-DIS-DVCS-DVMP}.

\begin{table*}  
\renewcommand{\arraystretch}{1.4}
\small
\caption{\label{tab:pars}
    The first two rows give the initial values of the parameters and limits on their ranges,
    if set. 
    The final fitted values 
    and their uncertainties (one standard deviation) of
    the best \model{NLO-DVCS-DVMP} model are given in the last two rows.
    The corresponding values of $\chinpts$ are given in the last column of Tab. \ref{tab:chis}.
    The parameters correspond to the scale  $\mu_{0}^2 = \qty{4}{\GeV^2}$.
}
\centering
\setlength{\tabcolsep}{3pt}
\renewcommand{\arraystretch}{1.2}
\begin{tabular}{cccccccccccc}
\hline\noalign{\smallskip}
& $N^{\Psea}$ & $\alpha_{0}^{\Psea}$ & $\alpha^{\prime}_{\Psea}$ & $m_{\Psea}^2$ & $s_{2}^{\Psea}$ & $s_{4}^{\Psea}$ & $\alpha_{0}^{\PG}$ & $\alpha^{\prime}_{\PG}$ & $m_{\PG}^2$ & $s_{2}^{\PG}$ & $s_{4}^{\PG}$ \\
 unit  & $$ & $1$ & $\unit{\GeV}^{-2}$ & $\unit{\GeV}^2$ & $1$ & $1$ & $1$ & $\unit{\GeV}^{-2}$ & $\unit{\GeV}^2$ & $1$ & $1$ \\
\hline
 initial  & 0.15 & 1.00 & 0.15 & 0.70 & -0.20 & 0.00 & 1.00 & 0.15 & 0.70 & 0.00 & 0.00 \\
 limits  &  &  & (0.0,1.0) & (0,3) & (-0.3,0.3) & (-0.1,0.1) &  & (0.0,1.0) & (0,3) & (-3.0,3.0) & (-1.0,1.0) \\
 final  & 0.168 & 1.128 & 0.125 & 0.412 & 0.280 & -0.044 & 1.099 & 0.000 & 0.145 & 2.958 & -0.951 \\
 uncert.  & 0.002 & 0.011 & 0.043 & 0.056 & 0.037 & 0.012 & 0.011 & 0.011 & 0.008 & 0.032 & 0.023 \\
\noalign{\smallskip}\hline
\end{tabular}
\renewcommand{\arraystretch}{1.}
\end{table*}

The initial values of the parameters are listed in the first row of Tab.
\ref{tab:pars}.
They were chosen very generically, using nevertheless to some
extent the knowledge of the corresponding Regge trajectories, so initial intercepts
are set to \num{1}, and initial slopes to \num{0.15} for both quarks and gluons. Also,
the first subleading quark partial wave normalization is set to a negative initial value
$s_{2}^{\Psea} = -0.2$ based on earlier LO DVCS studies \cite{Kumericki:2009uq}
which preferred it so. Some parameters are restricted as stated in the
second row of Tab. \ref{tab:pars}. The squares of masses $m_{a}^2$ and Regge
slopes $\alpha'^{a}$ are for physical reasons limited to positive values of the
order of one.
Negative values of $\alpha'^{a}$ enable solutions with discontinuous behavior
of the form factor as a function of $t$, which should be avoided.
Partial-wave normalizations of the quarks are constrained to achieve a
natural hierarchy $1 \gg s^{\Psea}_2 \gg s^{\Psea}_4$. Unfortunately a similar
constraint in the gluon sector prevented a good fit, so
gluon partial waves are left more flexible than quark ones.
Future studies should show whether the final relatively large values of
formally subleading gluon partial waves $s^{\PG}_2$ and $s^{\PG}_4$,
which were also observed in earlier studies that also included data from
experiments with a fixed target \cite{Kumericki:2015lhb},
are a signal of a problem with our model.
The resulting values of these and other parameters for the best of the six models,
\model{NLO-DVCS-DVMP}, are displayed in the third row of Tab. \ref{tab:pars},
and the uncertainties are given in the fourth row, where everything is the result of a standard least squares
fit with the software routine MINUIT \cite{James:1975dr,iminuit}.
The most correlated pairs of parameters for that model are
\begin{itemize}
     \item $\alpha'^{\Psea}$ and $m_{\Psea}^2$ with correlation \num{0.937} which
         shows that it is not possible to clearly distinguish the $t$ dependence associated
         with $x$ (``shrinkage'' effect, controlled by parameter $\alpha'$) from the
         residual $t$ dependence controlled by the parameter $m^2$,
     \item $s^{\Psea}_{2}$ and $s^{\Psea}_{4}$ with anticorrelation \num{-0.940},
         and $s^{\PG}_{2}$ and $s^{\PG}_{4}$ with anticorrelation \num{-0.931},
         suggesting that presently it is not possible to distinguish
         the second from the third SO(3) partial wave.
\end{itemize}
Regarding the goodness of the six fits, Tab. \ref{tab:chis} shows the
obtained values of $\chinpts$ where we observe as follows.
\begin{itemize}
     \item Models describe data well only if they are fitted to them
           (see Tab. \ref{tab:models}).
         For other data not included in the fitting, the obtained values $\chinpts$ are
         in the range \numrange{10}{30000}, and these cases are marked as $\gg{}1$ in Tab. \ref{tab:chis}.
     \item Only the NLO model \model{NLO-DVCS-DVMP} describes
         all the data well and therefore we consider it the best model presented in this paper.
     \item LO models can describe DIS and DVCS data well, as shown already in \cite{Kumericki:2007sa}, but they cannot describe DVMP well, with $\chinpts > 3$ 
         (although the total $\chinpts = 1.4$ of the 
         model \model{LO-DVCS-DVMP} for the total data set looks satisfactory).
\end{itemize}
In Tab. \ref{tab:chis}, we give the values for $\chi^2$ divided by the number of data
points and not, as is common, by the number of degrees of freedom. 
This is because for particular subsets of the total dataset,
it is not possible to distinguish the corresponding number of degrees of freedom. 
Let us give therefore
for the sake of completeness also $\chi^2/n_{\textup{d.o.f.}}$ for models fitted
to the maximal dataset DIS+DVCS+DVMP:
\begin{equation}
    \chi^2/n_{\textup{d.o.f.}}(\model{LO-DVCS-DVMP}) = 1.5 \;, \qquad \qquad
    \chi^2/n_{\textup{d.o.f.}}(\model{NLO-DVCS-DVMP}) = 1.2
\;.\end{equation}
We repeated the entire analysis on the data where L/T separation was performed
using the universal function $R(W, \Q^2)$, i.e., not using separate
$R(W, \Q^2)_{\textup{H1}}$ and $R(W, \Q^2)_{\textup{ZEUS}}$.
In that case, the conclusions are unchanged,
the plots are barely distinguishable, and the values $\chi^2/n_{\textup{d.o.f.}}$
also change only a little. For example, the value $\chi^2/n_{\textup{d.o.f.}}$
of the best model \model{NLO-DVCS-DVMP} becomes 1.3 instead of 1.2.

\begin{table*}  
\renewcommand{\arraystretch}{1.4}
\caption{\label{tab:chis}
    Values of $\chinpts$ for each LO or NLO model (columns) for the total DIS + DVCS + DVMP dataset and for subsets corresponding to different processes (rows).
(The values denoted by $\gg 1$ are greater than 10.)
}
\centering
\setlength{\tabcolsep}{8pt}
\renewcommand{\arraystretch}{1.2}
\begin{tabular}{ccccccccc}
\hline\noalign{\smallskip}
\multirow{2}{*}{Dataset} & \multirow{2}{*}{Refs.} & \multirow{2}{*}{$n_{\textnormal{pts}}$} &
  \multicolumn{3}{c}{\model{LO-}} & \multicolumn{3}{c}{\model{NLO-}}  \\
   & & & \model{DVCS} & \model{DVMP} & \model{DVCS-DVMP} & 
         \model{DVCS} & \model{DVMP} & \model{DVCS-DVMP} \\
\noalign{\smallskip}\hline
DIS & \cite{H1:1996zow} & 85 & 0.6 & 0.6 & 0.6 & 0.8 & 0.8 & 0.8 \\
DVCS & \cite{H1:2005gdw,H1:2009wnw,ZEUS:2003pwh,ZEUS:2008hcd} & 27 & 0.4 & $\gg{}1$ & 0.6 & 0.6 & $\gg{}1$ & 0.8 \\
DVMP & \cite{H1:2009cml,ZEUS:2007iet} & 45 & $\gg{}1$ & 3.1 & 3.3 & $\gg{}1$ & 1.5 & 1.8 \\
Total &  & 157 & $\gg{}1$ & $\gg{}1$ & 1.4 & 3.7 & $\gg{}1$ & 1.1 \\
\noalign{\smallskip}\hline
\end{tabular}
\renewcommand{\arraystretch}{1.}
\end{table*}

As stated in Sec. \ref{sec:gpdmodel}, instead of setting a forward limit of
our GPD model to be equal to PDF functions taken from one of the modern PDF
sets, we also fit models to DIS data, i.e, $F_2$ structure function whose
form in terms of the Mellin-Barnes integral is
given in \cite{Kumericki:2007sa}, Eq. (198).
The fit result is shown in Fig.
\ref{fig:H1_dis} and in the first line of Tab. \ref{tab:chis}, where it can be
seen that the description of these data is perfect.
This may seem a relatively trivial test of our model, but given the subtleties
when choosing the number of quark flavors and crossing the corresponding
thresholds, we believe that it has not been given enough attention in other attempts
to extract GPDs in the literature.
In Fig. \ref{fig:H1_dis}, we also present more precise combined H1 and ZEUS data, which
we have not utilized in our fits for reasons outlined in footnote \ref{fn:h115}. In the
illustrated region, where due to the small value of electron inelasticity the $F_2$ structure
function dominates the DIS cross-section, even this highly accurate data is reasonably described
by our models.

As can be seen in the second row of Tab. \ref{tab:chis} and in Fig. \ref{fig:H1ZEUS_dvcs}, the description of DVCS data is good in both LO and NLO. There are certain problems visible when describing ZEUS data in the
third panel of Fig. \ref{fig:H1ZEUS_dvcs}, but it is for a low value $\Q^2 =
\,\qty{3.2}{\GeV^2}$ and these data were not even included in the fit.

In Figs. \ref{fig:H1_dvmp} and \ref{fig:ZEUS_dvmp}, and in the third row of
Tab. \ref{tab:chis} results  are given related to one of the central
questions that we deal with in this work: Is it possible to describe the cross sections
for DVMP in the canonical collinear GPD approach? It can be seen immediately, from the problem
in the description of the $\Q^2$ dependence for higher values of $\Q^2 > \qty{30}{\GeV^2}$,
especially for the H1 data in Fig. \ref{fig:H1_dvmp} on the left,
and from the problem in the description of the dependence on $W$,
especially for the ZEUS data in Fig. \ref{fig:ZEUS_dvmp} on the right, and
from large values $\chinpts > 3$, that at the LO level the description is not good.
However, at NLO, the situation improves noticeably. True, the right
panels of Figs. \ref{fig:H1_dvmp} and \ref{fig:ZEUS_dvmp} suggest that the $W$ 
dependence of the NLO model is not steep enough (interestingly, for the LO models
it is too steep), but the description is mostly satisfactory, the behavior
for large values of $\Q^2$ is correct and $\chinpts$ values are acceptable.
It is worth focusing additionally on the problem of the $\Q^2$ dependence which
was often highlighted as the main problem of a purely collinear approach,
because in that approach, the canonical scaling of the longitudinal cross section
at a fixed value $\xb$ is of the form $\sim {1}/{\Q^6}$, cf. (\ref{eq:sigDVMPt}) and
(\ref{eq:HVconvolution}), whereas
experimental data on the total cross section %
reportedly suggest a milder dependence $\sim {1}/{\Q^4}$.
There, of course, the first problem is to distinguish the contributions of the purely longitudinal
cross section, which we dealt with in section \ref{sec:exp}.
If we accept the L/T separation function proposed there, in fact
the longitudinal component of the cross section also shows the behavior
$\sigDVMPL \sim {1}/{\Q^4}$; however, it should be noted that experimental
data for fixed $\xb$ are relatively scarce. They are shown in Fig. \ref{fig:Q2scaling}
where we see that within the experimental errors both \model{LO-DVMP} and \model{NLO-DVMP}
show satisfactory effective $\Q^2$ scaling in the measured range, which
is a consequence of perturbative $\Q^2$ logarithms. Only for much larger
values of $\Q^2$ the models obey the canonical $\sim {1}/{\Q^6}$ behavior.
We can better assess the quality of the description of the $\Q^2$ dependence of the cross section
if we study it for fixed $W$, for which
there is more abundant experimental information.
As discussed in section \ref{sec:exp}, the longitudinal cross section
for fixed $W$ shows the effective behavior $\sigDVMPL \propto \Q^{-5}$,
so in Fig. \ref{fig:Q2Wscaling} we show the rescaled $\Q^5 \sigDVMPL$ data compared
to four of our models fitted to DVMP measurements.
It can be seen that only NLO models can describe the behavior of $\sigDVMPL$ for large $\Q^2$,
with fixed $W$.

\subsection{Quark vs gluons at LO and NLO}
\label{sec:QvsG}

If we grant that the description of the experimental data presented in the previous section
is satisfying, let us see what we can learn about
the quark-gluon structure of the proton responsible for that description.
As the DIS and DVCS processes are related in the sense that they
use a photon probe that couples to the same flavor singlet
combination of quarks and gluons, separation of the quark and gluon structure
using only those two processes is necessarily indirect.
Adding the DVMP process to the analysis brings a new, meson probe, 
with, in principle, 
a different flavor structure, and at the same time a more direct 
approach to the gluonic GPD that already contributes 
to the LO amplitude, see Fig. \ref{fig:DVMPDVCS}.
Therefore, DVMP rightly stands out as the key to separating individual flavors
and gluonic GPD in the proton.

However, the influence of the gluon GPD on the DVCS cross section should not be underestimated 
either. This influence is reflected through two effects. 
The first of them is strong evolution
of gluons at high energies, which is already present in DIS observables
and is visible in Fig. \ref{fig:QvsGDIS}, where the contribution of gluons to the total
DIS $F_2(\xb, Q^2)$ at $\xb = 0.001$ is of the same sign and becomes equal to the
quark one already for $Q^2 \sim \qty{50}{\GeV^2}$.
Another effect is the significant contribution of gluons to the total NLO non-forward DVCS amplitude
of hard scattering. It is \emph{negative}, i.e. of the opposite sign to the quark contribution,
and despite the formal $\alphas/(2\pi)$ suppression is equally large,
which was already pointed out earlier \cite{Freund:2001rk,Diehl:2007zu}.
This can be clearly seen in the second panel of the Fig. \ref{fig:QvsGDVCS}, where already at
the initial scale $\mu_{0}^2 = \qty{4}{\GeV^2}$ a large negative contribution of gluons
would completely cancel out the quark LO contribution, so the NLO quark contribution
must be twice as large as at LO so that the final CFF $\ImH$ could be
in accordance with the experiment.
This second effect is not present in DIS, which can be clearly seen in Fig. \ref{fig:QvsGDIS}
where the displacement of gluons on the right NLO panel is positive and very small, so
that the LO and NLO cases are barely distinguishable.
It should be noted that what corresponds to the DVCS
CFF function $\mathcal{H}$ in deep-inelastic scattering is not the structure
function $F_2$ but $F_1$, as discussed in Sec. \ref{sec:lessons}.
Unlike $F_2$, where the NLO correction is positive and small, see
Fig.~\ref{fig:QvsGDIS}, for $F_1$, the relative correction is negative, therefore in sign
agreement with DVCS. It is also larger than for $F_2$, but still turns out to be
about four times smaller than for the CFF $\mathcal{H}$.
It is also important to highlight 
that the quark and gluon contributions in Figs. \ref{fig:QvsGDIS}-\ref{fig:QvsGDVMP}
were separated according to 
(\ref{eq:HHVcpweEV})
i.e., they reveal the influence of quarks and gluons present at the initial scale $\muO$.
Thus although there is no true gluon contribution for DVCS at LO,
the influence of initial gluons on the quark GPD, which they exhibit through evolution,
and thus on the CFF, is revealed. 
As can be seen in Fig. \ref{fig:QvsGDVMP},
gluons equally dominate the DVMP amplitude at both LO and NLO and the main
NLO effect is a relative suppression of the real part of the amplitude.
Some models of the longitudinal production of light vector mesons have predicted
that the gluon GPD alone is almost sufficient to describe the amplitude of this process
at high energy and our results approximately confirm this.

Such different behavior of GPDs and corresponding PDFs at LO and NLO can be
succinctly described using the so-called skewness ratio of GPDs and PDFs:
\begin{equation}
    r^{a}(\Q^2) = \frac{H^{a}(x, \xi=x, t=0, \Q^2)}{H^{a}(x, \xi=0, t=0, \Q^2)} \,,
 \label{eq:skewness}
\end{equation}
which for small values of $x$ practically does not depend on $x$.
To derive $x$-space GPDs from those modeled in $j$-space,
one must perform an inversion of the Mellin-Barnes integral. 
This process is discussed in detail in \cite{Mueller:2005ed, Muller:2014wxa}. 
An example of a GPD on the cross-over line is provided in \cite{Kumericki:2009uq}, 
specifically in Eq. (43).
In papers \cite{Shuvaev:1999ce,Martin:2008gqx} it was proposed that a GPD,
up to the dependence on $t$, be completely determined
by the value of the corresponding PDF and the conformal values of the skewness ratio:
\begin{equation}
    r^{\mathrm{\PSigma}}\approx 1.65\;,\qquad 
    r^{\mathrm{\PG}}\approx 1.03\;.
    \label{eq:shuvaev}
\end{equation}
Obviously, the real situation is more complex and this quantity varies a lot
from LO to NLO.  An additional variation is brought by a completely different
interplay of quarks and gluons in the DVMP amplitude.
We can gain additional insight by observing the skewness
ratio (\ref{eq:skewness})  of all six
models, shown in Fig. \ref{fig:skewness}.  The thin lines
show the three LO models and it can be seen that the quark and
also the gluon GPDs fitted to DVMP, DVCS or combined DVMP+DVCS data are
drastically different.  However, almost universal GPD functions emerge at NLO
for the combined DVCS+DVMP fit and for fits where DVCS or DVMP data are turned
off (thick lines). The NLO skewness ratios are consistently higher than conformal ones
(\ref{eq:shuvaev}) and are asymptotically slightly above the value of 2 for
quarks and about 1.5 for gluons.  We see that the gluon GPD from the combined NLO
DVCS+DVMP fit is closer to that from the DVMP fit, which is
probably a reflection of the fact that DVMP has more influence on the
extraction of gluonic GPD than DVCS.

Considering the diverse roles of quarks and gluons, successfully achieving a
comprehensive simultaneous description of all three observed processes using
unique universal GPD functions, exemplified by the \model{NLO-DVCS-DVMP} model,
stands as compelling validation of the presented theoretical framework.

\section{Conclusions}
\label{sec:concl}

In this work we have revisited the NLO corrections to DVMP.
The main ingredients and advantages of the CPaW formalism
\cite{Mueller:2005ed,Mueller:2005nz,Kumericki:2007sa,Mueller:2013caa} 
have been provided along with revised expressions for 
deeply virtual production of longitudinally
polarized vector mesons. 
The impact of NLO corrections has been analyzed through
global fits to DIS, DVCS and 
\dvvlpO data,
in particular, $\Prhozero$ meson data, at small-$\xb$.

We conclude that a reliable and consistent description of the longitudinal component of the
DVMP cross sections at high energies and for $\Q^2 > \,\qty{10}{\GeV^2}$
is possible in the standard collinear pQCD approach, only if
NLO corrections are included. A simultaneous description of DIS, DVCS and DVMP processes
becomes feasible at the NLO level, thus revealing the proton structure
described by universal GPD functions.

\subsection*{Code availability}
In the interest of open and reproducible research, the computer code used in
the production of numerical results and graphs for this paper is available at
\url{https://github.com/openhep/nloimpact23} and \url{https://gepard.phy.hr}.
\vspace*{2ex}


\subsection*{Acknowledgments}
We are grateful to D. M\"{u}ller for collaborations during which most of the
theoretical framework used was set up.
We thank V. Braun and P. Kroll for discussions.
This publication is supported by the Croatian Science Foundation project
IP-2019-04-9709, by QuantiXLie Centre of Excellence through the grant
KK.01.1.1.01.0004, and by the EU Horizon 2020 research and innovation program,
STRONG-2020 project, under grant agreement No 824093.


\small
\bibliographystyle{JHEP-2}                                                                                

\providecommand{\href}[2]{#2}\begingroup\raggedright\endgroup

\end{document}

%% file: wenneker.tex
\usepackage[english]{babel} 

\usepackage{microtype} 

\usepackage{amsmath,amsfonts,amsthm} 

\usepackage[svgnames]{xcolor} 

\usepackage[small, labelfont=bf, up, textfont=it]{caption} 

\usepackage{booktabs} 

\usepackage{lastpage} 

\usepackage{graphicx} 

\usepackage{enumitem} 
\setlist{noitemsep} 

\usepackage{sectsty} 
\allsectionsfont{\usefont{OT1}{phv}{b}{n}} 

\usepackage{multirow}
\usepackage{amssymb}
\usepackage{hepunits}
\usepackage{hepnames}
\usepackage{xpatch}
\makeatletter
\xpatchcmd\@HepConStyle
 {\edef\@upcode{\updefault}}
 {\ifdefined\shapedefault\edef\@upcode{\shapedefault}\else\edef\@upcode{\updefault}\fi}
 {}{}
\makeatother
\usepackage[pagebackref=false,colorlinks=true,citecolor=blue,filecolor=blue,%
linkcolor=blue,anchorcolor=blue,urlcolor=blue]{hyperref}
\usepackage[colorinlistoftodos,tickmarkheight=5pt]{todonotes}
\newcounter{mycomment}

\newcounter{counttodoC}

\newcounter{counttodoTODO}

\newcounter{counttodoQ}

\usepackage{geometry} 

\usepackage{setspace}
\usepackage[T1]{fontenc} 
\usepackage[utf8]{inputenc} 

\usepackage{XCharter} 

\usepackage{fancyhdr} 
\fancypagestyle{firstpage}{ 
	\fancyhf{}
}

\newcommand{\authorstyle}[1]{{\large\usefont{OT1}{phv}{b}{n}\color{DarkRed}#1}} 

\newcommand{\institution}[1]{{\footnotesize\usefont{OT1}{phv}{m}{sl}\color{Black}#1}} 

\usepackage{titling} 

\newcommand{\HorRule}{\noindent\color{DarkGoldenrod}\rule{\linewidth}{1pt}} 

\pretitle{
	\vspace{-30pt} 
	\HorRule\vspace{10pt} 
	\fontsize{32}{36}\usefont{OT1}{phv}{b}{n}\selectfont 
	\color{DarkRed} 
}

\posttitle{\par\vskip 15pt} 

\preauthor{} 

\postauthor{ 
	\vspace{10pt} 
	\par\HorRule 
	\vspace{20pt} 
}

\usepackage{lettrine} 
\usepackage{fix-cm}	

\newcommand{\initial}[1]{ 
	\lettrine[lines=3,findent=4pt,nindent=0pt]{
		\color{DarkGoldenrod}
		{#1}
	}{}%
}

\usepackage{xstring} 

\newcommand{\lettrineabstract}[1]{
	\StrLeft{#1}{1}[\firstletter] 
	\initial{\firstletter}\textbf{\StrGobbleLeft{#1}{1}} 
}





%% file: nloimpact.bbl
\begin{thebibliography}{100}

\bibitem{Mueller:1998fv}
D.~Müller, D.~Robaschik, B.~Geyer, F.~M. Dittes and J.~Hořejší, {\it Wave
  functions, evolution equations and evolution kernels from light-ray operators
  of {QCD}},  {\em Fortschr. Phys.} {\bf 42} (1994) 101
  [\href{http://arXiv.org/abs/hep-ph/9812448}{{\tt hep-ph/9812448}}].

\bibitem{Radyushkin:1996nd}
A.~V. Radyushkin, {\it Scaling limit of deeply virtual {C}ompton scattering},
  {\em Phys. Lett.} {\bf B380} (1996) 417--425
  [\href{http://arXiv.org/abs/hep-ph/9604317}{{\tt hep-ph/9604317}}].

\bibitem{Ji:1996nm}
X.-D. Ji, {\it Deeply-virtual compton scattering},  {\em Phys. Rev.} {\bf D55}
  (1997) 7114--7125 [\href{http://arXiv.org/abs/hep-ph/9609381}{{\tt
  hep-ph/9609381}}].

\bibitem{Ji:1996ek}
X.-D. Ji, {\it Gauge invariant decomposition of nucleon spin},  {\em Phys. Rev.
  Lett.} {\bf 78} (1997) 610--613
  [\href{http://arXiv.org/abs/hep-ph/9603249}{{\tt hep-ph/9603249}}].

\bibitem{Burkardt:2000za}
M.~Burkardt, {\it Impact parameter dependent parton distributions and off-
  forward parton distributions for $\zeta \to 0$},  {\em Phys. Rev.} {\bf D62}
  (2000) 071503 [\href{http://arXiv.org/abs/hep-ph/0005108}{{\tt
  hep-ph/0005108}}]. Erratum-ibid.D66:119903,2002.

\bibitem{Polyakov:2002yz}
M.~V. Polyakov, {\it {Generalized parton distributions and strong forces inside
  nucleons and nuclei}},  {\em Phys. Lett.} {\bf B555} (2003) 57--62
  [\href{http://arXiv.org/abs/hep-ph/0210165}{{\tt hep-ph/0210165}}].

\bibitem{Accardi:2012qut}
A.~Accardi {\em et.~al.}, {\it {Electron Ion Collider: The Next QCD Frontier}},
   {\em Eur. Phys. J.} {\bf A52} (2016), no.~9 268
  [\href{http://arXiv.org/abs/1212.1701}{{\tt 1212.1701}}].

\bibitem{Anderle:2021wcy}
D.~P. Anderle {\em et.~al.}, {\it {Electron-ion collider in China}},  {\em
  Front. Phys. (Beijing)} {\bf 16} (2021), no.~6 64701
  [\href{http://arXiv.org/abs/2102.09222}{{\tt 2102.09222}}].

\bibitem{dHose:2016mda}
N.~d'Hose, S.~Niccolai and A.~Rostomyan, {\it {Experimental overview of Deeply
  Virtual Compton Scattering}},  {\em Eur. Phys. J.} {\bf A52} (2016), no.~6
  151.

\bibitem{Kumericki:2016ehc}
K.~Kumerički, S.~Liuti and H.~Moutarde, {\it {GPD phenomenology and DVCS
  fitting}},  {\em Eur. Phys. J.} {\bf A52} (2016), no.~6 157
  [\href{http://arXiv.org/abs/1602.02763}{{\tt 1602.02763}}].

\bibitem{Favart:2015umi}
L.~Favart, M.~Guidal, T.~Horn and P.~Kroll, {\it {Deeply Virtual Meson
  Production on the nucleon}},  {\em Eur. Phys. J. A} {\bf 52} (2016), no.~6
  158 [\href{http://arXiv.org/abs/1511.04535}{{\tt 1511.04535}}].

\bibitem{Berger:2001xd}
E.~R. Berger, M.~Diehl and B.~Pire, {\it {Time-like Compton scattering:
  Exclusive photoproduction of lepton pairs}},  {\em Eur. Phys. J. C} {\bf 23}
  (2002) 675--689 [\href{http://arXiv.org/abs/hep-ph/0110062}{{\tt
  hep-ph/0110062}}].

\bibitem{Pire:2011st}
B.~Pire, L.~Szymanowski and J.~Wagner, {\it {NLO corrections to timelike,
  spacelike and double deeply virtual Compton scattering}},  {\em Phys. Rev. D}
  {\bf 83} (2011) 034009 [\href{http://arXiv.org/abs/1101.0555}{{\tt
  1101.0555}}].

\bibitem{Boer:2015fwa}
M.~Bo\"er, M.~Guidal and M.~Vanderhaeghen, {\it {Timelike Compton scattering
  off the proton and generalized parton distributions}},  {\em Eur. Phys. J. A}
  {\bf 51} (2015), no.~8 103.

\bibitem{Grocholski:2019pqj}
O.~Grocholski, H.~Moutarde, B.~Pire, P.~Sznajder and J.~Wagner, {\it
  {Data-driven study of timelike Compton scattering}},  {\em Eur. Phys. J. C}
  {\bf 80} (2020), no.~2 171 [\href{http://arXiv.org/abs/1912.09853}{{\tt
  1912.09853}}].

\bibitem{Belitsky:2002tf}
A.~V. Belitsky and D.~Mueller, {\it {Exclusive electroproduction of lepton
  pairs as a probe of nucleon structure}},  {\em Phys. Rev. Lett.} {\bf 90}
  (2003) 022001 [\href{http://arXiv.org/abs/hep-ph/0210313}{{\tt
  hep-ph/0210313}}].

\bibitem{Guidal:2002kt}
M.~Guidal and M.~Vanderhaeghen, {\it {Double deeply virtual Compton scattering
  off the nucleon}},  {\em Phys. Rev. Lett.} {\bf 90} (2003) 012001
  [\href{http://arXiv.org/abs/hep-ph/0208275}{{\tt hep-ph/0208275}}].

\bibitem{Deja:2023ahc}
K.~Deja, V.~Martinez-Fernandez, B.~Pire, P.~Sznajder and J.~Wagner, {\it
  {Phenomenology of double deeply virtual Compton scattering in the era of new
  experiments}},  \href{http://arXiv.org/abs/2303.13668}{{\tt 2303.13668}}.

\bibitem{Ivanov:2002jj}
D.~Y. Ivanov, B.~Pire, L.~Szymanowski and O.~V. Teryaev, {\it {Probing chiral
  odd GPD's in diffractive electroproduction of two vector mesons}},  {\em
  Phys. Lett. B} {\bf 550} (2002) 65--76
  [\href{http://arXiv.org/abs/hep-ph/0209300}{{\tt hep-ph/0209300}}].

\bibitem{Boussarie:2016qop}
R.~Boussarie, B.~Pire, L.~Szymanowski and S.~Wallon, {\it {Exclusive
  photoproduction of a $\gamma\,\rho$ pair with a large invariant mass}},  {\em
  JHEP} {\bf 02} (2017) 054 [\href{http://arXiv.org/abs/1609.03830}{{\tt
  1609.03830}}]. [Erratum: JHEP 10, 029 (2018)].

\bibitem{Pedrak:2017cpp}
A.~Pedrak, B.~Pire, L.~Szymanowski and J.~Wagner, {\it {Hard photoproduction of
  a diphoton with a large invariant mass}},  {\em Phys. Rev. D} {\bf 96}
  (2017), no.~7 074008 [\href{http://arXiv.org/abs/1708.01043}{{\tt
  1708.01043}}]. [Erratum: Phys.Rev.D 100, 039901 (2019)].

\bibitem{Siddikov:2022bku}
M.~Siddikov and I.~Schmidt, {\it {Exclusive production of quarkonia pairs in
  collinear factorization framework}},  {\em Phys. Rev. D} {\bf 107} (2023),
  no.~3 034037 [\href{http://arXiv.org/abs/2212.14019}{{\tt 2212.14019}}].

\bibitem{Duplancic:2022ffo}
G.~Duplančić, S.~Nabeebaccus, K.~Passek-Kumerički, B.~Pire, L.~Szymanowski
  and S.~Wallon, {\it {Accessing chiral-even quark generalised parton
  distributions in the exclusive photoproduction of a
  \ensuremath{\gamma}\ensuremath{\pi}$^{±}$ pair with large invariant mass in
  both fixed-target and collider experiments}},  {\em JHEP} {\bf 03} (2023) 241
  [\href{http://arXiv.org/abs/2212.00655}{{\tt 2212.00655}}].

\bibitem{Qiu:2022pla}
J.-W. Qiu and Z.~Yu, {\it {Single diffractive hard exclusive processes for the
  study of generalized parton distributions}},  {\em Phys. Rev. D} {\bf 107}
  (2023), no.~1 014007 [\href{http://arXiv.org/abs/2210.07995}{{\tt
  2210.07995}}].

\bibitem{Diehl:2003ny}
M.~Diehl, {\it Generalized parton distributions},  {\em Phys. Rept.} {\bf 388}
  (2003) 41--277 [\href{http://arXiv.org/abs/hep-ph/0307382}{{\tt
  hep-ph/0307382}}].

\bibitem{Diehl:2013xca}
M.~Diehl and P.~Kroll, {\it {Nucleon form factors, generalized parton
  distributions and quark angular momentum}},  {\em Eur.Phys.J.} {\bf C73}
  (2013), no.~4 2397 [\href{http://arXiv.org/abs/1302.4604}{{\tt 1302.4604}}].

\bibitem{Radyushkin:1998rt}
A.~V. Radyushkin, {\it {Nonforward parton densities and soft mechanism for
  form-factors and wide angle Compton scattering in QCD}},  {\em Phys. Rev. D}
  {\bf 58} (1998) 114008 [\href{http://arXiv.org/abs/hep-ph/9803316}{{\tt
  hep-ph/9803316}}].

\bibitem{Diehl:1998kh}
M.~Diehl, T.~Feldmann, R.~Jakob and P.~Kroll, {\it {Linking parton
  distributions to form-factors and Compton scattering}},  {\em Eur. Phys. J.
  C} {\bf 8} (1999) 409--434 [\href{http://arXiv.org/abs/hep-ph/9811253}{{\tt
  hep-ph/9811253}}].

\bibitem{Huang:2000kd}
H.~W. Huang and P.~Kroll, {\it {Large momentum transfer electroproduction of
  mesons}},  {\em Eur. Phys. J. C} {\bf 17} (2000) 423--435
  [\href{http://arXiv.org/abs/hep-ph/0005318}{{\tt hep-ph/0005318}}].

\bibitem{Kroll:2021ecb}
P.~Kroll and K.~Passek-Kumeri\v{c}ki, {\it {Wide-angle photo- and
  electroproduction of pions to twist-3 accuracy}},  {\em Phys. Rev. D} {\bf
  104} (2021), no.~5 054040 [\href{http://arXiv.org/abs/2107.04544}{{\tt
  2107.04544}}].

\bibitem{Collins:1999un}
J.~C. Collins and M.~Diehl, {\it {Transversity distribution does not contribute
  to hard exclusive electroproduction of mesons}},  {\em Phys. Rev. D} {\bf 61}
  (2000) 114015 [\href{http://arXiv.org/abs/hep-ph/9907498}{{\tt
  hep-ph/9907498}}].

\bibitem{Goloskokov:2009ia}
S.~V. Goloskokov and P.~Kroll, {\it {An Attempt to understand exclusive pi+
  electroproduction}},  {\em Eur.Phys.J.} {\bf C65} (2010) 137--151
  [\href{http://arXiv.org/abs/0906.0460}{{\tt 0906.0460}}].

\bibitem{Goloskokov:2011rd}
S.~V. Goloskokov and P.~Kroll, {\it {Transversity in hard exclusive
  electroproduction of pseudoscalar mesons}},  {\em Eur. Phys. J. A} {\bf 47}
  (2011) 112 [\href{http://arXiv.org/abs/1106.4897}{{\tt 1106.4897}}].

\bibitem{Ji:1997nk}
X.-D. Ji and J.~Osborne, {\it {One-loop QCD corrections to deeply-virtual
  Compton scattering: The parton helicity-independent case}},  {\em Phys. Rev.}
  {\bf D57} (1998) 1337--1340 [\href{http://arXiv.org/abs/hep-ph/9707254}{{\tt
  hep-ph/9707254}}].

\bibitem{Belitsky:1997rh}
A.~V. Belitsky and D.~Müller, {\it {Predictions from conformal algebra for the
  deeply virtual Compton scattering}},  {\em Phys. Lett.} {\bf B417} (1998)
  129--140 [\href{http://arXiv.org/abs/hep-ph/9709379}{{\tt hep-ph/9709379}}].

\bibitem{Mankiewicz:1997bk}
L.~Mankiewicz, G.~Piller, E.~Stein, M.~Vanttinen and T.~Weigl, {\it {NLO
  corrections to deeply-virtual Compton scattering}},  {\em Phys. Lett.} {\bf
  B425} (1998) 186--192 [\href{http://arXiv.org/abs/hep-ph/9712251}{{\tt
  hep-ph/9712251}}].

\bibitem{Ji:1998xh}
X.-D. Ji and J.~Osborne, {\it {One-loop corrections and all order factorization
  in deeply virtual Compton scattering}},  {\em Phys. Rev.} {\bf D58} (1998)
  094018 [\href{http://arXiv.org/abs/hep-ph/9801260}{{\tt hep-ph/9801260}}].

\bibitem{Belitsky:2001nq}
A.~V. Belitsky and D.~Mueller, {\it {Hard exclusive meson production at
  next-to-leading order}},  {\em Phys. Lett. B} {\bf 513} (2001) 349--360
  [\href{http://arXiv.org/abs/hep-ph/0105046}{{\tt hep-ph/0105046}}].

\bibitem{Ivanov:2004zv}
D.~Y. Ivanov, L.~Szymanowski and G.~Krasnikov, {\it {Vector meson
  electroproduction at next-to-leading order}},  {\em JETP Lett.} {\bf 80}
  (2004) 226--230 [\href{http://arXiv.org/abs/hep-ph/0407207}{{\tt
  hep-ph/0407207}}]. [Erratum: JETP Lett. 101, 844 (2015)].

\bibitem{Duplancic:2016bge}
G.~Duplan\v{c}i\'c, D.~M\"uller and K.~Passek-Kumeri\v{c}ki, {\it
  {Next-to-leading order corrections to deeply virtual production of
  pseudoscalar mesons}},  {\em Phys. Lett. B} {\bf 771} (2017) 603--610
  [\href{http://arXiv.org/abs/1612.01937}{{\tt 1612.01937}}].

\bibitem{Braun:2020yib}
V.~M. Braun, A.~N. Manashov, S.~Moch and J.~Schoenleber, {\it {Two-loop
  coefficient function for DVCS: vector contributions}},  {\em JHEP} {\bf 09}
  (2020) 117 [\href{http://arXiv.org/abs/2007.06348}{{\tt 2007.06348}}].
  [Erratum: JHEP 02, 115 (2022)].

\bibitem{Braun:2021grd}
V.~M. Braun, A.~N. Manashov, S.~Moch and J.~Schoenleber, {\it {Axial-vector
  contributions in two-photon reactions: Pion transition form factor and
  deeply-virtual Compton scattering at NNLO in QCD}},  {\em Phys. Rev. D} {\bf
  104} (2021), no.~9 094007 [\href{http://arXiv.org/abs/2106.01437}{{\tt
  2106.01437}}].

\bibitem{Ji:2023xzk}
Y.~Ji and J.~Schoenleber, {\it {Two-loop coefficient functions in deeply
  virtual Compton scattering: flavor-singlet axial-vector and transversity
  case}},  \href{http://arXiv.org/abs/2310.05724}{{\tt 2310.05724}}.

\bibitem{Braun:2011zr}
V.~M. Braun and A.~N. Manashov, {\it {Kinematic power corrections in
  off-forward hard reactions}},  {\em Phys. Rev. Lett.} {\bf 107} (2011) 202001
  [\href{http://arXiv.org/abs/1108.2394}{{\tt 1108.2394}}].

\bibitem{Braun:2011dg}
V.~M. Braun and A.~N. Manashov, {\it {Operator product expansion in QCD in
  off-forward kinematics: Separation of kinematic and dynamical
  contributions}},  {\em JHEP} {\bf 01} (2012) 085
  [\href{http://arXiv.org/abs/1111.6765}{{\tt 1111.6765}}].

\bibitem{Braun:2014sta}
V.~M. Braun, A.~N. Manashov, D.~M\"uller and B.~M. Pirnay, {\it {Deeply Virtual
  Compton Scattering to the twist-four accuracy: Impact of finite-$t$ and
  target mass corrections}},  {\em Phys. Rev. D} {\bf 89} (2014), no.~7 074022
  [\href{http://arXiv.org/abs/1401.7621}{{\tt 1401.7621}}].

\bibitem{Guo:2021gru}
Y.~Guo, X.~Ji and K.~Shiells, {\it {Higher-Order Kinematical Effects in Deeply
  Virtual Compton Scattering}},  {\em JHEP} {\bf 12} (2021) 103
  [\href{http://arXiv.org/abs/2109.10373}{{\tt 2109.10373}}].

\bibitem{Braun:2022qly}
V.~M. Braun, Y.~Ji and A.~N. Manashov, {\it {Next-to-leading-power kinematic
  corrections to DVCS: a scalar target}},  {\em JHEP} {\bf 01} (2023) 078
  [\href{http://arXiv.org/abs/2211.04902}{{\tt 2211.04902}}].

\bibitem{Kroll:2012sm}
P.~Kroll, H.~Moutarde and F.~Sabatie, {\it {From hard exclusive meson
  electroproduction to deeply virtual Compton scattering}},  {\em Eur.Phys.J.}
  {\bf C73} (2013), no.~1 2278 [\href{http://arXiv.org/abs/1210.6975}{{\tt
  1210.6975}}].

\bibitem{Goloskokov:2005sd}
S.~V. Goloskokov and P.~Kroll, {\it {Vector meson electroproduction at small
  Bjorken-x and generalized parton distributions}},  {\em Eur. Phys. J.} {\bf
  C42} (2005) 281--301 [\href{http://arXiv.org/abs/hep-ph/0501242}{{\tt
  hep-ph/0501242}}].

\bibitem{Goloskokov:2007nt}
S.~V. Goloskokov and P.~Kroll, {\it {The Role of the quark and gluon GPDs in
  hard vector-meson electroproduction}},  {\em Eur.Phys.J.} {\bf C53} (2008)
  367--384 [\href{http://arXiv.org/abs/0708.3569}{{\tt 0708.3569}}].

\bibitem{Meskauskas:2011aa}
M.~Meškauskas and D.~Müller, {\it {A Fresh Look at Exclusive
  Electroproduction of Light Vector Mesons}},  {\em Eur. Phys. J. C} {\bf 74}
  (2014), no.~2 2719 [\href{http://arXiv.org/abs/1112.2597}{{\tt 1112.2597}}].

\bibitem{Lautenschlager:2013uya}
T.~Lautenschlager, D.~Müller and A.~Schäfer, {\it {Global analysis of
  generalized parton distributions --collider kinematics--}},
  \href{http://arXiv.org/abs/1312.5493}{{\tt 1312.5493}}. (unpublished).

\bibitem{Lautenschlager:2015mxt}
T.~Lautenschlager, {\em {Towards a global estimate for generalized parton
  distributions}}.
\newblock PhD thesis, Regensburg U., 2015.

\bibitem{Mueller:2005ed}
D.~Müller and A.~Schäfer, {\it Complex conformal spin partial wave expansion
  of generalized parton distributions and distribution amplitudes},  {\em Nucl.
  Phys.} {\bf B739} (2006) 1--59
  [\href{http://arXiv.org/abs/hep-ph/0509204}{{\tt hep-ph/0509204}}].

\bibitem{Mueller:2005nz}
D.~Müller, {\it Next-to-next-to leading order corrections to deeply virtual
  compton scattering: The non-singlet case},  {\em Phys. Lett. B} {\bf 634}
  (2006) 227--234 [\href{http://arXiv.org/abs/hep-ph/0510109}{{\tt
  hep-ph/0510109}}].

\bibitem{Kumericki:2007sa}
K.~Kumerički, D.~Müller and K.~Passek-Kumerički, {\it Towards a fitting
  procedure for deeply virtual {C}ompton scattering at next-to-leading order
  and beyond},  {\em Nucl. Phys.} {\bf B794} (2008) 244--323
  [\href{http://arXiv.org/abs/hep-ph/0703179}{{\tt hep-ph/0703179}}].

\bibitem{Mueller:2013caa}
D.~Müller, T.~Lautenschlager, K.~Passek-Kumerički and A.~Schäfer, {\it
  {Towards a fitting procedure to deeply virtual meson production -- the
  next-to-leading order case --}},  {\em Nucl. Phys. B} {\bf 884} (2014)
  438--546 [\href{http://arXiv.org/abs/1310.5394}{{\tt 1310.5394}}].

\bibitem{Lepage:1980fj}
G.~P. Lepage and S.~J. Brodsky, {\it {Exclusive Processes in Perturbative
  Quantum Chromodynamics}},  {\em Phys. Rev. D} {\bf 22} (1980) 2157.

\bibitem{Efremov:1979qk}
A.~V. Efremov and A.~V. Radyushkin, {\it {Factorization and Asymptotical
  Behavior of Pion Form-Factor in QCD}},  {\em Phys. Lett. B} {\bf 94} (1980)
  245--250.

\bibitem{Brodsky:1994kf}
S.~J. Brodsky, L.~Frankfurt, J.~F. Gunion, A.~H. Mueller and M.~Strikman, {\it
  {Diffractive leptoproduction of vector mesons in QCD}},  {\em Phys. Rev. D}
  {\bf 50} (1994) 3134--3144 [\href{http://arXiv.org/abs/hep-ph/9402283}{{\tt
  hep-ph/9402283}}].

\bibitem{Martin:1996bp}
A.~D. Martin, M.~G. Ryskin and T.~Teubner, {\it {The QCD description of
  diffractive rho meson electroproduction}},  {\em Phys. Rev. D} {\bf 55}
  (1997) 4329--4337 [\href{http://arXiv.org/abs/hep-ph/9609448}{{\tt
  hep-ph/9609448}}].

\bibitem{Diehl:2007hd}
M.~Diehl and W.~Kugler, {\it {Next-to-leading order corrections in exclusive
  meson production}},  {\em Eur. Phys. J. C} {\bf 52} (2007) 933--966
  [\href{http://arXiv.org/abs/0708.1121}{{\tt 0708.1121}}].

\bibitem{Ivanov:2007je}
D.~Y. Ivanov, {\it {Exclusive vector meson electroproduction}},  in {\em {12th
  International Conference on Elastic and Diffractive Scattering: Forward
  Physics and QCD}}, pp.~26--32, 12, 2007.
\newblock \href{http://arXiv.org/abs/0712.3193}{{\tt 0712.3193}}.

\bibitem{Collins:1996fb}
J.~C. Collins, L.~Frankfurt and M.~Strikman, {\it {Factorization for hard
  exclusive electroproduction of mesons in QCD}},  {\em Phys. Rev. D} {\bf 56}
  (1997) 2982--3006 [\href{http://arXiv.org/abs/hep-ph/9611433}{{\tt
  hep-ph/9611433}}].

\bibitem{Collins:1998be}
J.~C. Collins and A.~Freund, {\it {Proof of factorization for deeply virtual
  Compton scattering in QCD}},  {\em Phys. Rev. D} {\bf 59} (1999) 074009
  [\href{http://arXiv.org/abs/hep-ph/9801262}{{\tt hep-ph/9801262}}].

\bibitem{Kumericki:2006xx}
K.~Kumerički, D.~Müller, K.~Passek-Kumerički and A.~Schäfer, {\it Deeply
  virtual compton scattering beyond next-to-leading order: The flavor singlet
  case},  {\em Phys. Lett.} {\bf B648} (2007) 186--194
  [\href{http://arXiv.org/abs/hep-ph/0605237}{{\tt hep-ph/0605237}}].

\bibitem{Ivanov:2004vd}
D.~Y. Ivanov, A.~Schafer, L.~Szymanowski and G.~Krasnikov, {\it {Exclusive
  photoproduction of a heavy vector meson in QCD}},  {\em Eur. Phys. J. C} {\bf
  34} (2004), no.~3 297--316 [\href{http://arXiv.org/abs/hep-ph/0401131}{{\tt
  hep-ph/0401131}}]. [Erratum: Eur.Phys.J.C 75, 75 (2015)].

\bibitem{Jones:2015nna}
S.~P. Jones, A.~D. Martin, M.~G. Ryskin and T.~Teubner, {\it {Exclusive
  $J/\psi$ and $\Upsilon$ photoproduction and the low $x$ gluon}},  {\em J.
  Phys. G} {\bf 43} (2016), no.~3 035002
  [\href{http://arXiv.org/abs/1507.06942}{{\tt 1507.06942}}].

\bibitem{Vinnikov:2006xw}
A.~V. Vinnikov, {\it {Code for prompt numerical computation of the leading
  order GPD evolution}},  \href{http://arXiv.org/abs/hep-ph/0604248}{{\tt
  hep-ph/0604248}}.

\bibitem{Bertone:2022frx}
V.~Bertone, H.~Dutrieux, C.~Mezrag, J.~M. Morgado and H.~Moutarde, {\it
  {Revisiting evolution equations for generalised parton distributions}},  {\em
  Eur. Phys. J. C} {\bf 82} (2022), no.~10 888
  [\href{http://arXiv.org/abs/2206.01412}{{\tt 2206.01412}}].

\bibitem{Berthou:2015oaw}
B.~Berthou {\em et.~al.}, {\it {PARTONS: PARtonic Tomography Of Nucleon
  Software}: {A computing framework for the phenomenology of Generalized Parton
  Distributions}},  {\em Eur. Phys. J. C} {\bf 78} (2018), no.~6 478
  [\href{http://arXiv.org/abs/1512.06174}{{\tt 1512.06174}}].

\bibitem{Freund:2001bf}
A.~Freund and M.~F. McDermott, {\it {Next-to-leading order evolution of
  generalized parton distributions for DESY HERA and HERMES}},  {\em Phys. Rev.
  D} {\bf 65} (2002) 056012 [\href{http://arXiv.org/abs/hep-ph/0106115}{{\tt
  hep-ph/0106115}}]. [Erratum: Phys.Rev.D 66, 079903 (2002)].

\bibitem{Braun:2017cih}
V.~M. Braun, A.~N. Manashov, S.~Moch and M.~Strohmaier, {\it {Three-loop
  evolution equation for flavor-nonsinglet operators in off-forward
  kinematics}},  {\em JHEP} {\bf 06} (2017) 037
  [\href{http://arXiv.org/abs/1703.09532}{{\tt 1703.09532}}].

\bibitem{Braun:2022bpn}
V.~M. Braun, Y.~Ji and J.~Schoenleber, {\it {Deeply Virtual Compton Scattering
  at Next-to-Next-to-Leading Order}},  {\em Phys. Rev. Lett.} {\bf 129} (2022),
  no.~17 172001 [\href{http://arXiv.org/abs/2207.06818}{{\tt 2207.06818}}].

\bibitem{Gross:1974cs}
D.~J. Gross and F.~Wilczek, {\it {Asymptotically Free Gauge Theories. 2.}},
  {\em Phys. Rev. D} {\bf 9} (1974) 980--993.

\bibitem{Georgi:1974wnj}
H.~Georgi and H.~D. Politzer, {\it {Electroproduction scaling in an
  asymptotically free theory of strong interactions}},  {\em Phys. Rev. D} {\bf
  9} (1974) 416--420.

\bibitem{Floratos:1981hs}
E.~G. Floratos, C.~Kounnas and R.~Lacaze, {\it {Higher Order QCD Effects in
  Inclusive Annihilation and Deep Inelastic Scattering}},  {\em Nucl. Phys.}
  {\bf B192} (1981) 417.

\bibitem{Belitsky:1998uk}
A.~V. Belitsky, D.~Müller, L.~Niedermeier and A.~Schäfer, {\it Evolution of
  non-forward parton distributions in next-to- leading order: Singlet sector},
  {\em Nucl. Phys.} {\bf B546} (1999) 279--298
  [\href{http://arXiv.org/abs/hep-ph/9810275}{{\tt hep-ph/9810275}}].

\bibitem{Melic:2001wt}
B.~Melić, B.~Nižić and K.~Passek, {\it {A Note on the factorization scale
  independence of the PQCD predictions for exclusive processes}},  {\em Eur.
  Phys. J. C} {\bf 36} (2004) 453--458
  [\href{http://arXiv.org/abs/hep-ph/0107311}{{\tt hep-ph/0107311}}].

\bibitem{Kumericki:2009uq}
K.~Kumerički and D.~Müller, {\it {Deeply virtual Compton scattering at small
  $x_{\rm B}$ and the access to the GPD H}},  {\em Nucl. Phys.} {\bf B841}
  (2010) 1--58 [\href{http://arXiv.org/abs/0904.0458}{{\tt 0904.0458}}].

\bibitem{Muller:2014wxa}
D.~Müller, M.~V. Polyakov and K.~M. Semenov-Tian-Shansky, {\it {Dual
  parametrization of generalized parton distributions in two equivalent
  representations}},  {\em JHEP} {\bf 1503} (2015) 052
  [\href{http://arXiv.org/abs/1412.4165}{{\tt 1412.4165}}].

\bibitem{Moutarde:2018kwr}
H.~Moutarde, P.~Sznajder and J.~Wagner, {\it {Border and skewness functions
  from a leading order fit to DVCS data}},  {\em Eur. Phys. J.} {\bf C78}
  (2018), no.~11 890 [\href{http://arXiv.org/abs/1807.07620}{{\tt
  1807.07620}}].

\bibitem{Kriesten:2021sqc}
B.~Kriesten, P.~Velie, E.~Yeats, F.~Yepez~Lopez and S.~Liuti, {\it
  {Parametrization of quark and gluon generalized parton distributions in a
  dynamical framework}},  {\em Phys. Rev. D} {\bf 105} (2022), no.~5 056022
  [\href{http://arXiv.org/abs/2101.01826}{{\tt 2101.01826}}].

\bibitem{Guo:2022upw}
Y.~Guo, X.~Ji and K.~Shiells, {\it {Generalized parton distributions through
  universal moment parameterization: zero skewness case}},  {\em JHEP} {\bf 09}
  (2022) 215 [\href{http://arXiv.org/abs/2207.05768}{{\tt 2207.05768}}].

\bibitem{Guo:2023ahv}
Y.~Guo, X.~Ji, M.~G. Santiago, K.~Shiells and J.~Yang, {\it {Generalized parton
  distributions through universal moment parameterization: non-zero skewness
  case}},  {\em JHEP} {\bf 05} (2023) 150
  [\href{http://arXiv.org/abs/2302.07279}{{\tt 2302.07279}}].

\bibitem{Braun:2016wnx}
V.~M. Braun {\em et.~al.}, {\it {The \ensuremath{\rho}-meson light-cone
  distribution amplitudes from lattice QCD}},  {\em JHEP} {\bf 04} (2017) 082
  [\href{http://arXiv.org/abs/1612.02955}{{\tt 1612.02955}}].

\bibitem{H1:2009cml}
{\bf H1} Collaboration, F.~D. Aaron {\em et.~al.}, {\it {Diffractive
  Electroproduction of $\rho$ and $\phi$ Mesons at HERA}},  {\em JHEP} {\bf 05}
  (2010) 032 [\href{http://arXiv.org/abs/0910.5831}{{\tt 0910.5831}}].

\bibitem{ZEUS:2007iet}
{\bf ZEUS} Collaboration, S.~Chekanov {\em et.~al.}, {\it {Exclusive $\rho^0$
  production in deep inelastic scattering at HERA}},  {\em PMC Phys.} {\bf A1}
  (2007) 6 [\href{http://arXiv.org/abs/0708.1478}{{\tt 0708.1478}}].

\bibitem{H1:1996zow}
{\bf H1} Collaboration, S.~Aid {\em et.~al.}, {\it {A Measurement and QCD
  analysis of the Proton Structure Function $F_2(x, Q^2)$ at HERA}},  {\em
  Nucl. Phys. B} {\bf 470} (1996) 3--40
  [\href{http://arXiv.org/abs/hep-ex/9603004}{{\tt hep-ex/9603004}}].

\bibitem{H1:2015ubc}
{\bf H1, ZEUS} Collaboration, H.~Abramowicz {\em et.~al.}, {\it {Combination of
  measurements of inclusive deep inelastic ${e^{\pm }p}$ scattering cross
  sections and QCD analysis of HERA data}},  {\em Eur. Phys. J. C} {\bf 75}
  (2015), no.~12 580 [\href{http://arXiv.org/abs/1506.06042}{{\tt
  1506.06042}}].

\bibitem{H1:2005gdw}
{\bf H1} Collaboration, A.~Aktas {\em et.~al.}, {\it {Measurement of deeply
  virtual Compton scattering at HERA}},  {\em Eur. Phys. J.} {\bf C44} (2005)
  1--11 [\href{http://arXiv.org/abs/hep-ex/0505061}{{\tt hep-ex/0505061}}].

\bibitem{H1:2009wnw}
{\bf H1} Collaboration, F.~Aaron {\em et.~al.}, {\it {Deeply Virtual Compton
  Scattering and its Beam Charge Asymmetry in $e^{\pm}p$ Collisions at HERA}},
  {\em Phys.Lett.} {\bf B681} (2009) 391--399
  [\href{http://arXiv.org/abs/0907.5289}{{\tt 0907.5289}}].

\bibitem{ZEUS:2003pwh}
{\bf ZEUS} Collaboration, S.~Chekanov {\em et.~al.}, {\it {Measurement of
  deeply virtual Compton scattering at HERA}},  {\em Phys. Lett.} {\bf B573}
  (2003) 46--62 [\href{http://arXiv.org/abs/hep-ex/0305028}{{\tt
  hep-ex/0305028}}].

\bibitem{ZEUS:2008hcd}
{\bf ZEUS} Collaboration, S.~Chekanov {\em et.~al.}, {\it {A Measurement of the
  $Q^2$, W and t dependences of deeply virtual Compton scattering at HERA}},
  {\em JHEP} {\bf 0905} (2009) 108 [\href{http://arXiv.org/abs/0812.2517}{{\tt
  0812.2517}}].

\bibitem{Alexeev:2022gkb}
G.~D. Alexeev {\em et.~al.}, {\it {Spin Density Matrix Elements in Exclusive
  $\rho ^0$ Meson Muoproduction}},  \href{http://arXiv.org/abs/2210.16932}{{\tt
  2210.16932}}.

\bibitem{COMPASS:2020zre}
{\bf COMPASS} Collaboration, M.~G. Alexeev {\em et.~al.}, {\it {Spin density
  matrix elements in exclusive $\omega $ meson muoproduction}},  {\em Eur.
  Phys. J. C} {\bf 81} (2021), no.~2 126
  [\href{http://arXiv.org/abs/2009.03271}{{\tt 2009.03271}}].

\bibitem{Kumericki:2015lhb}
K.~Kumerički and D.~Müller, {\it {Description and interpretation of DVCS
  measurements}},  {\em EPJ Web Conf.} {\bf 112} (2016) 01012
  [\href{http://arXiv.org/abs/1512.09014}{{\tt 1512.09014}}].

\bibitem{James:1975dr}
F.~James and M.~Roos, {\it {\sc Minuit}: A system for function minimization and
  analysis of the parameter errors and correlations},  {\em Comput. Phys.
  Commun.} {\bf 10} (1975) 343--367.

\bibitem{iminuit}
H.~Dembinski, P.~Ongmongkolkul {\em et.~al.}, {\it scikit-hep/iminuit},  {\em
  doi:10.5281/zenodo.3949207} (Dec, 2020).

\bibitem{Freund:2001rk}
A.~Freund and M.~F. McDermott, {\it {A Next-to-leading order QCD analysis of
  deeply virtual Compton scattering amplitudes}},  {\em Phys. Rev. D} {\bf 65}
  (2002) 074008 [\href{http://arXiv.org/abs/hep-ph/0106319}{{\tt
  hep-ph/0106319}}].

\bibitem{Diehl:2007zu}
M.~Diehl and W.~Kugler, {\it {Some numerical studies of the evolution of
  generalized parton distributions}},  {\em Phys. Lett.} {\bf B660} (2008)
  202--211 [\href{http://arXiv.org/abs/0711.2184}{{\tt 0711.2184}}].

\bibitem{Shuvaev:1999ce}
A.~G. Shuvaev, K.~J. Golec-Biernat, A.~D. Martin and M.~G. Ryskin, {\it
  {Off-diagonal distributions fixed by diagonal partons at small x and xi}},
  {\em Phys. Rev.} {\bf D60} (1999) 014015
  [\href{http://arXiv.org/abs/hep-ph/9902410}{{\tt hep-ph/9902410}}].

\bibitem{Martin:2008gqx}
A.~D. Martin, C.~Nockles, M.~G. Ryskin, A.~G. Shuvaev and T.~Teubner, {\it
  {Generalised parton distributions at small x}},  {\em Eur. Phys. J. C} {\bf
  63} (2009) 57--67 [\href{http://arXiv.org/abs/0812.3558}{{\tt 0812.3558}}].

\end{thebibliography}
